%% file: N-agent-MFG-limit-SIFIN-Re.tex
\documentclass[a4,11pt]{article}
\usepackage{graphicx} 
\usepackage{float}
\usepackage{amsmath,amsthm}
\usepackage{amsfonts}
\usepackage{latexsym}
\usepackage{amssymb}
\usepackage{setspace}
\usepackage{comment}
\usepackage{cases}
\usepackage{color}
\makeatletter
 
  \@addtoreset{equation}{section}
 \makeatother

\begin{document}
\title{{Strong Convergence to the Mean-Field Limit \\of A Finite Agent Equilibrium}
~\footnote{
To appear in SIAM Journal on Financial Mathematics.
Previously titled as {\it A Finite Agent Equilibrium in an Incomplete Market and its Strong Convergence to the Mean-Field Limit}.
}}

\author{Masaaki Fujii\footnote{Quantitative Finance Course, Graduate School of Economics, The University of Tokyo. }, \quad
Akihiko Takahashi\footnote{Quantitative Finance Course, Graduate School of Economics, The University of Tokyo. }
}
\date{ 
First version:  21 October, 2020\\
This version: 10 December, 2021
}
\maketitle


\input{Fmacro-2015.tex}

\def\calf{{\cal F}}
\def\wt{\widetilde}
\def\mbb{\mathbb}
\def\ol{\overline}
\def\ul{\underline}
\def\sign{{\rm{sign}}}
\def\wh{\widehat}
\def\vr{\varrho}
\def\p{\prime}
\def\pp{^\prime}
\def\ep{\epsilon}
\def\vep{\varepsilon}
\def\del{\delta}
\def\Del{\Delta}
\def\enu{\mathfrak{n}}

\def\mg{\mathfrak}

\def\ch{\check}
\def\ak{\alpha^{(k)}}
\def\as{\alpha^{(s)}}
\def\Ito{It\^o}

\def\hatl{\widehat{\lambda}}
\def\part{\partial}
\def\ubar{\underbar}
\def\ul{\underline}
\def\ol{\overline}
\def\ha{\widehat{\alpha}}
\def\hc{\widehat{c}}
\def\vp{\varpi}
\def\nn{\nonumber}
\def\be{\begin{equation}}
\def\ee{\end{equation}}
\def\bea{\begin{eqnarray}}
\def\eea{\end{eqnarray}}
\def\beas{\begin{eqnarray*}}
\def\eeas{\end{eqnarray*}}
\def\tbf{\textbf}
\def\bg{\boldsymbol}

\def\bull{$\bullet~$}

\newcommand{\Slash}[1]{{\ooalign{\hfil/\hfil\crcr$#1$}}}
\vspace{-5mm}
\begin{abstract}
We study an equilibrium-based continuous asset pricing problem for the securities market.
In the previous work~\cite{Fujii-Takahashi}, we have shown that a certain price process,
which is  given by the solution to a forward backward stochastic differential equation of conditional McKean-Vlasov type, 
asymptotically clears the market in the large population limit.
In the current work, under suitable conditions, we show the existence of a finite agent equilibrium and 
its strong convergence to the corresponding mean-field limit given in \cite{Fujii-Takahashi}.
As an important byproduct, we get the direct estimate on the difference of the equilibrium price between 
the two markets; the one consisting of heterogeneous agents of finite population size and the other of
homogeneous agents of infinite population size.

\end{abstract}


\section{Introduction}
In many of the applications of financial mathematics, such as optimal trading  and derivatives pricing,  
the securities price processes are typically assumed exogenously, for example, 
by the Black-Scholes, Heston or SABR models.
The parameters used in the processes are somehow calibrated so that they reproduce, at least approximately, 
the important properties observed in the market.
This is  by far  the dominant approach being adopted by practitioners due to its flexibility and simplicity for implementation.
However, within this framework, we cannot ask how and why such price processes appear in the market.
In particular,  the relation between the price processes and the characters
of the market participants are just left unquestioned.
The central theme of the current paper is to determine the price processes endogenously 
in terms of the behaviors of rational financial firms by requiring a rather obvious but a very important condition: 
the demand and supply of the securities
must always be balanced, which is the so-called the {\it market clearing condition}.
This is the problem of equilibrium price formation.
We are going to derive the price processes based on the agents' preferences (i.e. cost functions)
and investigate what happens in the large population limit.

The problem of equilibrium asset price formation has been one 
of the central issues in financial economics for a long time.
Its intrinsic difficulty comes from the stochastic differential games among many agents.
Recent developments of  Mean Field Game (MFG)  theory have opened a new interesting 
approach to multi-agent problems. Since the publication of the pioneering works by Lasry \& Lions \cite{Lions-1, Lions-2, Lions-3}
and Huang, Malhame \& Caines~\cite{Caines-Huang, Caines-Huang-1, Caines-Huang-2, Caines-Huang-3},
mean field game theory has been one of the most active research topics in various fields.
The strength of the mean field approach resides in the fact that it decomposes a difficult 
problem of a stochastic differential game into a tractable individual optimization problem and 
an additional fixed-point problem in the large population limit. It has been proved that the 
solution to the mean-field game equilibrium gives an $\ep$-Nash equilibrium 
for the corresponding game of finite homogeneous agents.
For interested readers, there are excellent monographs such as \cite{Bensoussan-mono, Gomes-eco, Gomes-reg, Kolokoltsov-mono}
for analytic approach and \cite{Carmona-Delarue-1, Carmona-Delarue-2} for probabilistic approach based on 
forward-backward stochastic differential equations (FBSDEs) of McKean-Vlasov type. 
See also \cite{Lacker-Delarue, Djete, Djete-Possamai-1, Lacker-1, Lacker-2, Lacker-3} for another approach using the concept of relaxed controls,
which does not produce any equation characterizing the equilibrium solution but can significantly weaken
the regularity assumptions we need.  

Since the mean-field game theory has been developed for the analysis of the Nash equilibrium, examples of its direct applications
to the market clearing equilibrium are very hard to find. In the majority of  works, 
certain phenomenological approaches are taken. One popular approach is to suppose
that the asset price process is decomposed into two parts, one is 
an exogenous process which is independent of the agents' action,
and the other representing the market friction (i.e. price impact) which is often assumed to be proportional to the average trading speed of the agents.
Another approach is to impose the market clearing 
condition but the demand of the asset is assumed to be given by an exogenous function of  price
without  considering the optimization problem among the agents.
See \cite{Matoussi, Djehiche-E, Evangelista, Feron,Fu-Horst-1, Fu-Horst-2,Fu,Gueant-Oil,Lehalle}
as interesting applications to, optimal trading, optimal liquidation, optimal oil production,  and price formation in electricity 
markets etc., using the phenomenological approaches explained above.
Although this approach makes the setup nicely fit to the Nash game, the market clearing equilibrium cannot be investigated anymore.
In fact,  the relation between the price processes and the trading activities are just assumed exogenously.
Notable exceptions dealing with the market clearing equilibrium in the large population limit 
can be found in \cite{Gomes-Saude, Firoozi} for electricity markets,
where the price process becomes deterministic due to the absence of the common noise.

In the previous paper \cite{Fujii-Takahashi}, we have investigated
the problem of equilibrium price formation in an incomplete securities market.
Each financial firm (agent) was assumed to minimize its cost via continuous-time trading with the securities exchange 
while facing the systemic and idiosyncratic noises as well as the stochastic order-flows from its over-the-counter  clients.
In contrast to \cite{Gomes-Saude, Firoozi}, the analysis was carried out in the presence of common noise.
The price process of the $n$ securities  $(\vp_t)_{t\in [0,T]}$ is only 
required to be square integrable and progressively measurable.
The adopted cost function is a natural generalization of those used in optimal liquidation problems.
The running as well as the terminal costs depend not only on the position size of the securities but also on the 
equilibrium price $\vp$ which is to be determined endogenously.
We found that the solution to a certain FBSDE of conditional McKean-Vlasov type 
gives an approximate of the equilibrium price process which clears the market asymptotically in the large population limit.

The crucial variables are the (optimal) trading speed $\wh{\alpha}^i_t$ of each agent $i$, $1\leq i\leq N$.
The quantity $\wh{\alpha}_t^i dt$ denotes the number of shares of the securities bought (or sold if negative) at the exchange
within the time interval $[t,t+dt]$ by the $i$th agent.
We found in \cite{Fujii-Takahashi} that a certain price process  $(\vp_t)_{t\in[0,T]}$ gives rise to the optimal trading strategies  $(\wh{\alpha}^i)_{i\in \mbb{N}}$ among the agents
which satisfy
\be
\lim_{N\rightarrow \infty}\frac{1}{N}\sum_{i=1}^N \wh{\alpha}_t^i=0, \quad dt\otimes d\mbb{P}-a.e. 
\label{eq-asymptotic-clearing}
\ee
This is the asymptotic market clearing in the {\it average} sense, i.e. the excess demand per capita converges to zero.
However, two important question remains unanswered.
Firstly,  the existence of the market clearing equilibrium among finite number of agents remains unknown.
Secondly,  although the price process $(\vp_t)_{t\in[0,T]}$ clears the market asymptotically, it does not
directly tell how close it is to the equilibrium securities price process in the finite population market (if it exists).
In the current paper, we answer these questions. 
Our first contribution is the proof of existence of the market clearing equilibrium in the finite population market. 
By relaxing the information assumption for each agent, we are able to characterize
the equilibrium among the $N$ agents by a system of $N$ fully-coupled FBSDEs.
The existence of a unique solution is proved by exploiting the convexity 
as well as monotone conditions of the coefficient functions.
Although the mathematical technique is the standard one~\cite{Peng-Wu}, to the best of the authors' knowledge, 
this is the first application of the method for proving the existence of equilibrium in an incomplete market
of finite population size.
In contrast to the existing literature, most of which adopt the exponential-type utility function with respect to the terminal wealth,
the proposed method allows us to handle general functional forms for  cost (or utility) functions as long as they satisfy 
a certain set of convexity and monotonicity conditions.
Moreover,  the existing works usually suppose {\it a priori} that the price process to have a simple diffusion form with  constant volatility
and focus solely on finding an appropriate drift term, while we only require the minimal assumption of square integrability. See, for examples, \cite{Jarrow-Book, Zitkovic-1, Xing-Zitkovic, Weston-Zitkovic} and references therein. 

Our second (and main) contribution is  to build a direct bridge connecting the equilibrium among the finite number of agents 
and its large population limit by showing the strong convergence  of the system of FBSDEs to 
that of conditional McKean-Vlasov type found in \cite{Fujii-Takahashi}.
Since the convergence involves the backward components of the FBSDEs, it can be called the {\it backward propagation of chaos}.
Note that,  this is a very rare example where one can prove the convergence of the equilibrium of 
finite number of agents to the corresponding mean-field limit,  except the cases of explicitly solvable linear quadratic setups. 
Although one can find related results on backward propagation of chaos in the recent work~\cite{Lauriere},
let us emphasize that the proof of convergence based on the monotonicity conditions for an arbitrary time interval
was given in the first time in \cite{FT-finite-agent} (Oct. 2020), which is the first preprint version 
in arXiv of our current work. We shall give more details on this point in Remark~\ref{remark-tangpi}.
In addition, as an important byproduct,  we  obtain the stability relation for the market clearing price 
between the two markets, the one is the finite population market and the other is its
large population limit. Using the stability property of the FBSDEs, we also give direct estimate on the difference between 
the equilibrium price among the heterogeneous agents and that for the homogeneous mean-field limit in terms of the 
difference of coefficients and the size of population.
Finally, the convergence to the mean-field limit reveals a role of the securities market 
as an efficient filter which removes the idiosyncratic noises from the equilibrium price process.
As the studies of the price formation with common noise, let us also refer the very recent publications~\cite{Gomes-etal-1, Gomes-etal-2}.
In particular, \cite{Gomes-etal-2} investigates the price formation of a single commodity with a random supply by making use of the 
method of calculus of variations, where assumptions on the state dynamics as well as the cost functions are also different from ours. 

The organization of the paper is as follows:
In Section 2, the notations used in the paper are explained.
In Section 3, the first major result regarding the existence of the unique equilibrium among the finite number of
agents is given  (Theorems~\ref{th-N-equilibrium}  and \ref{th-fbsde-N-existence}).
Section 4 is devoted to prove the strong convergence of the finite-agent equilibrium to its
mean field limit (Theorem~\ref{th-strong-convergence}), which is the second major result of the paper. The stability result between the equilibrium price for the finite heterogeneous 
agents and the mean field limit of homogeneous agents is also discussed. Concluding remarks are given in Section 5.
\section{Notations}
\label{sec-notation}

We use the same notations adopted in the work~\cite{Fujii-Takahashi}.
We introduce (N+1) complete probability spaces:
\bea
(\ol{\Omega}^0, \ol{\calf}^0,\ol{\mbb{P}}^0) \quad {\rm{and}} \quad (\ol{\Omega}^i,\ol{\calf}^i,\ol{\mbb{P}}^i)_{i=1}^N~,\nn
\eea
endowed with filtrations $\ol{\mbb{F}}^i:=(\ol{\calf}_t^i)_{t\geq 0}$, $i\in\{0,\cdots, N\}$.
Here, $\ol{\mbb{F}}^0$ is the completion of the 
filtration generated by $d^0$-dimensional Brownian motion $W^0$ (hence right-continuous)
and, for each $i\in\{1,\cdots, N\}$,  $\ol{\mbb{F}}^i$ is the complete and right-continuous augmentation of the filtration
generated by $d$-dimensional Brownian motions $W^i$
as well as a $W^i$-independent $n$-dimensional  square-integrable random variables $(\xi^i)$. 
Basically, the quantities indexed by zero are relevant for the noise and information common 
for all the agents.
We also introduce the product probability spaces
\be
\Omega^i=\ol{\Omega}^0\times \ol{\Omega}^i, \quad \calf^i, \quad \mbb{F}^i=(\calf_t^i)_{t\geq 0}, \quad \mbb{P}^i~, i\in\{1,\cdots, N\}\nn
\ee
where $(\calf^i,\mbb{P}^i)$ is the completion of $(\ol{\calf}^0\otimes \ol{\calf}^i, \ol{\mbb{P}}^0\otimes \ol{\mbb{P}}^i)$
and $\mbb{F}^i$ is the complete and right-continuous augmentation of $(\ol{\calf}_t^0\otimes \ol{\calf}_t^i)_{t\geq 0}$.
In the same way, we  define  the probability space $(\Omega,\calf,\mbb{P})$ endowed with $\mbb{F}=(\calf_t)_{t\geq 0}$
as a product of $(\ol{\Omega}^i,\ol{\calf}^i,\ol{\mbb{P}}^i;\ol{\mbb{F}}^i)_{i=0}^N$. 

For discussing the large population limit, we denote by $(\ol{\Omega}^\infty, \ol{\calf}^\infty, \ol{\mbb{P}}^\infty;\ol{\mbb{F}}^\infty)$
the product of $(\ol{\Omega}^i,\ol{\calf}^i,\ol{\mbb{P}}^i;\ol{\mbb{F}}^i)_{i=0}^\infty$,
and  also by $(\Omega^\infty,\calf^\infty, \mbb{P}^\infty;\mbb{F}^\infty)$ the product of $(\ol{\Omega}^0,\ol{\calf}^0, \ol{\mbb{P}}^0;\ol{\mbb{F}}^0)$
and $(\ol{\Omega}^\infty, \ol{\calf}^\infty, \ol{\mbb{P}}^\infty;\ol{\mbb{F}}^\infty)$,
constructed in the same fashion as above.
Unless otherwise stated, every probability space is complete and the every filtration satisfies the
usual conditions by standard completion and augmentation.

\vspace{2mm}
Throughout the work, the symbol $L$ denotes a given positive constant, 
the symbol $C$ a general positive constant which may change line by line. 
For a given constant $T>0$ and any measurable space $(\Omega,\calg)$
with the filtration $\mbb{G}:=(\calg_t)_{t\geq 0}$, we use the following notations for frequently encountered spaces:\\
\bull $\mbb{L}^2(\calg; \mbb{R}^d)$ denotes the set of $\mbb{R}^d$-valued $\calg$-measurable 
square integrable random variables.\\
\bull $\mbb{S}^2(\mbb{G};\mbb{R}^d)$ is the set of $\mbb{R}^d$-valued $\mbb{G}$-adapted continuous processes $X$ satisfying
\bea
||X||_{\mbb{S}^2}:=\mbb{E}\bigl[\sup_{t\in[0,T]}|X_t|^2\bigr]^\frac{1}{2}<\infty~. \nn
\eea
\bull $\mbb{H}^2(\mbb{G};\mbb{R}^d)$ is the set of $\mbb{R}^d$-valued $\mbb{G}$-progressively measurable processes $Z$ satisfying
\bea
||Z||_{\mbb{H}^2}:=\mbb{E}\Bigl[\Bigl(\int_0^T |Z_t|^2 dt\Bigr)\Bigr]^\frac{1}{2}<\infty~. \nn
\eea
\bull $\call(X)$ denotes the law of a random variable $X$.\\
$\bullet~\calp(\mbb{R}^d)$ is the set of probability measures on $(\mbb{R}^d,\calb(\mbb{R}^d))$. \\
$\bullet~\calp_p(\mbb{R}^d)$ with $p\geq 1$ is the subset of $\calp(\mbb{R}^d)$ with finite $p$-th moment; i.e.,
the set of $\mu\in \calp(\mbb{R}^d)$ satisfying
\bea
M_p(\mu):=\Bigl(\int_{\mbb{R}^d}|x|^p \mu(dx)\Bigr)^\frac{1}{p}<\infty~.\nn
\eea
We always assign $\calp_p(\mbb{R}^d)$ with $(p\geq 1)$ the $p$-Wasserstein distance $W_p$,
which makes  $\calp_p(\mbb{R}^d)$ a complete separable metric space.
It is defined by, for any $\mu, \nu\in \calp_p(\mbb{R}^d)$, 
\bea
W_p(\mu,\nu):={\inf}_{\pi\in\Pi_p(\mu,\nu)}\Bigl[\Bigl(\int_{\mbb{R}^d\times \mbb{R}^d} |x-y|^p \pi(dx,dy)\Bigr)^\frac{1}{p}\Bigr]
\label{def-W}
\eea
where $\Pi_p(\mu,\nu)$ denotes the set of probability measures in $\calp_p (\mbb{R}^d\times \mbb{R}^d)$
with marginals $\mu$ and $\nu$. For more details,  see \cite[Chapter 5]{Carmona-Delarue-1}. \\
\bull $m(\mu)$ denotes the expectation with respect to $\mu\in \calp(\mbb{R}^d)$, i.e.
\bea
m(\mu):=\int_{\mbb{R}^d} x\mu(dx). \nn
\eea
\bull For any $N$ variables $(x^i)_{i=1}^N$, we write its empirical mean as
\be
\mg{m}((x^i)):=\frac{1}{N}\sum_{i=1}^N x^i.\nn
\ee
We frequently omit the arguments such as $(\mbb{G},\mbb{R}^d)$ in the above definitions when there is no confusion 
from the context.

\section{Equilibrium in the finite population market}
\label{sec-N-equilibrium}
Our first goal is to prove the existence of the unique market clearing equilibrium for a stylized 
model of securities market and its characterization by the system of FBSDEs.
Although the securities market we are going to study is essentially the same as the one used in the previous work~\cite[Section 3]{Fujii-Takahashi},
we shall give the details for the readers' convenience.

\subsection{Description of the problem}
\label{subsec-model}
We consider the equilibrium-based pricing problem 
of $n$ types of  securities labeled by $k$,  $1\leq k\leq n$, which are 
continuously traded via the securities exchange in the presence of a large number of rational financial firms (agents)
indexed by $i$,  $1\leq i\leq N$.  Throughout this section, we work on the probability space $(\Omega,\calf,\mbb{P})$
and the expectation is taken under $\mbb{P}$.

Every agent is supposed to have many small individual clients
who can trade the securities only with the agent via the over-the-counter (OTC) market and have no direct access to the
exchange.\footnote{In fact,  only credit-worthy registered financial firms are allowed to directly participate in the securities exchange.
The individual investors and non-financial firms can trade the securities with these registered firms 
playing the role of financial intermediaries. This is called the over-the-counter (OTC) market.}
We denote the market price process of the $n$ securities by an $\mbb{R}^n$-valued process $(\vp_t)_{t\in[0,T]}$,
the detailed mathematical properties of which are to be discussed later. Here, $(\vp_t)^k$ denotes the market price of the $k$th security at time $t$.
In our model, the state process $(X_t^i)_{t\in[0,T]}$ of each agent $i$,  $1\leq i\leq N$, is given by the time evolution of 
his/her position size in the $n$ securities.  For example, let us suppose that the $k$th security is an equity of a certain firm.
Then $(X_t^i)^k$ denotes the number of shares of the equity possessed by the $i$th agent at time $t$.
If it is negative, it means that the agent is taking the {\it short} position.
Each agent $i$, $1\leq i\leq N$,  controls the trading speed of the securities $(\alpha_t^i)_{t\in[0,T]}$ via the  exchange
within some space of admissible strategies $\mbb{A}$.
More precisely,  $(\alpha^i_t)^k dt$, $1\leq k\leq n$, denotes the number of shares of the $k$th security
bought (or sold if negative) within the time interval $[t,t+dt]$ by the $i$th agent.
In addition to the trading via the exchange, the position size of each agent is affected 
by his/her market making via the OTC market with individual clients.
Although,   in the real market,  each financial firm dynamically controls bid-offer spreads in order to earn trading fees
and to affect the order flows  from his/her clients in a favorable manner to his/her profit, we treat, in this work, the order flows via the OTC market exogenous
and concentrate on the optimal trading problem via the securities exchange for simplicity.
We denote by $(c_t^0)_{t\geq 0}\in \mbb{H}^2(\ol{\mbb{F}}^0;\mbb{R}^n)$ with $c_T^0\in \mbb{L}^2(\ol{\calf}_T^0;\mbb{R}^n)$
the cash flows from the securities or the market news commonly available to all the agents, 
while by $(c_t^i)_{t\geq 0}\in \mbb{H}^2(\ol{\mbb{F}}^i;\mbb{R}^n)$ with $c_T^i\in \mbb{L}^2(\ol{\calf}_T^i;\mbb{R}^n)$ 
some independent factors and news affecting only 
on the agent $i$.\footnote{The dimensions of $c^0$ and $c^i$ are chosen to be $n$ only for the notational simplicity. One can assign 
any fixed dimensions for them so that they can represent any factors that affect the agents' cost functions. }
Moreover,  we assume that $(c_t^i)_{t\geq 0}$  have the common law for all $1\leq i\leq N$.

Let us introduce the measurable functions, $l_i:[0,T]\times (\mbb{R}^n)^3\rightarrow \mbb{R}^n$,
$\sigma_i^0:[0,T]\times (\mbb{R}^n)^3\rightarrow \mbb{R}^{n\times d_0}$ and $\sigma_i:[0,T]\times (\mbb{R}^n)^3\rightarrow \mbb{R}^{n\times d}$,
for $1\leq i\leq N$.
Using them, we now express the state dynamics (i.e. position size) of each agent $i$, $1\leq i\leq N$, by
\bea
dX_t^i=\bigl(\a_t^i+l_i(t,\vp_t,c_t^0, c_t^i)\bigr)dt+\sigma_i^0(t,\vp_t,c_t^0, c_t^i)dW_t^0+\sigma_i(t,\vp_t,c_t^0,c_t^i)dW_t^i \nn
\eea
with $X_0^i=\xi^i \in \mbb{L}^2(\ol{\calf}_0^i;\mbb{R}^n)$. $\xi^i$ denotes the initial position size of the $i$th agent
and is assumed to be independently and identically distributed (i.i.d.) among $1\leq i\leq N$.
In addition to $\alpha_t^i dt$ representing the change due to the direct trading via the exchange,
there also exist contributions from the order flows via the OTC market: $l_i(t,\vp_t, c_t^0,c_t^i)dt$ and $\bigl(\sigma_i^0(t,\vp_t,c_t^0,c_t^i)dW_t^0$ 
$,\sigma_i(t,\vp_t,c_t^0,c_t^i)dW_t^i\bigr)$ denote their finite and infinite variation parts, respectively.
We naturally expect that these order flows are dependent on the price of the securities, common as well as idiosyncratic informations.
Suppose,  for example, $(l_i)^k(t,\vp_t,c_t^0,c_t^i)<0$.  This means that the clients of the $i$th agent  
are buying the $k$th security from the agent via the OTC market with the net speed $|(l_i)^k(t,\vp_t,c_t^0,c_t^i)|$ at time $t$. 
The two infinite variation terms represent the noise in the order flows.
Note that, in addition to the random initial states $(\xi^i)_{i=1}^N$, we have $d_0$-dimensional common noise $W^0$
and $N$ $d$-dimensional idiosyncratic noises $(W^i)_{i=1}^N$. Since we impose no restriction 
on the size among $(n, d_0, d, N)$, we have an incomplete securities market in general.

Under such an environment, each agent tries to minimize his/her cost by controlling the trading speed $\alpha^i:=(\alpha_t^i)_{t\in[0,T]}$.
We suppose that the problem for each agent $1\leq i\leq N$ is given by
\be
\inf_{\a^i \in \mbb{A}}J^i (\a^i)~,
\label{agent-P}
\ee
where the cost functional $J^i(\cdot)$ will be specified later.
The space of admissible controls  $\mbb{A}$ is assumed to be common for every agent 
and is given by $\mbb{A}:=\mbb{H}^2(\mbb{F};\mbb{R}^n)$, i.e. the space of $\mbb{F}$-progressively measurable processes $\alpha$
satisfying
\be
\mbb{E}\int_0^T |\alpha_t|^2 dt<\infty. \nn
\ee
\begin{remark}
\label{remark-info}
The above definition of the space of admissible strategy $\mbb{A}=\mbb{H}^2(\mbb{F};\mbb{R}^n)$
implies that each agent knows, in addition to the common information $\ol{\mbb{F}}^0$, 
all the idiosyncratic information $(\ol{\mbb{F}}^i)_{i=1}^N$.
In other words, we assume the so-called {\it the perfect information}.
To the best of the authors' knowledge, the same information assumption is used in the existing literature
dealing with the market equilibrium with finite population. Ideally, we would like to restrict the information set available to each agent $i$ to the filtration 
$(\sigma\{\vp_s: s\leq t\}\vee \calf_t^i)_{t\geq 0}$. 
Interestingly, we shall see in later sections that the above idealistic
situation is actually realized in the large population limit.
\end{remark}

\begin{definition}[market clearing condition]
The market clearing condition is defined by
\be
\sum_{i=1}^N \alpha_t^i=0,  \quad dt\otimes d\mbb{P}{\text{-}}a.e., \nn
\ee
i.e. the demand and supply of each security  always balance 
among the $N$ agents.
\end{definition}

\begin{definition}[market clearing equlibrium]
If there exists an optimal solution $\wh{\alpha}^i$ to $(\ref{agent-P})$ for every agent $1\leq i\leq N$, and 
if the set of $(\wh{\alpha}^i)_{i=1}^N$ satisfies the market clearing condition, we call the solution the 
market clearing equilibrium. 
\end{definition}

For each agent $1\leq i \leq N$, let us introduce the following cost functions; $f_i:[0,T]\times (\mbb{R}^n)^5\rightarrow \mbb{R}$, 
$g_i:(\mbb{R}^n)^4\rightarrow \mbb{R}$,
$\ol{f}_i:[0,T]\times (\mbb{R}^n)^4\rightarrow \mbb{R}$ and $\ol{g}_i:(\mbb{R}^n)^3\rightarrow \mbb{R}$, which 
are Borel measurable and supposed to have the following form:
\be
\begin{split}
&f_i(t,x,\a,\vp,c^0,c):=\langle \vp,\a\rangle+\frac{1}{2}\langle \a,\L\a\rangle+\ol{f}_i(t,x,\vp,c^0,c),  \\
&g_i(x,\vp,c^0,c):=-b \langle \vp,x\rangle+\ol{g}_i(x,c^0,c)~. 
\end{split}
\label{eq-cost-functions}
\ee 
The associated cost functional is defined by
\bea
J^i( \a^i):=\mbb{E}\Bigl[\int_0^T f_i(t,X_t^i,\a_t^i,\vp_t,c_t^0, c_t^i)dt+g_i(X_T^i,\vp_T,c_T^0,c_T^i)\Bigr]~. \nn
\eea
In the above expression, $f_i$ and $g_i$ denote the running and the terminal costs, respectively.
Let us explain the economic meaning of each term.
By buying (or selling if negative) with speed $\alpha_t$, each agent pays (or receives if negative)
$\langle \alpha_t, \vp_t\rangle dt$ amount of cash in the time interval $[t,t+dt]$.
In addition to this direct cost,  we suppose that each agent has to pay the service fees 
to the securities exchange $\frac{1}{2}\langle \alpha_t, \Lambda \alpha_t\rangle dt$ where
$\Lambda$ is an $n\times n$ positive definite matrix. 
These costs are represented by the first two terms of the function $f_i$.
The first term of $g_i$ denotes
the mark-to-market value at the closing time with some discount factor $b<1$.\footnote{We shall see that the condition $b<1$
is necessary to obtain well-defined terminal condition for the equilibrium.}
The above three terms are assumed to be common across the agents 
since there is no strong motivation to suppose otherwise.

The remaining terms represented by functions $\ol{f}_i$ and $\ol{g}_i$
can be used to distinguish various characters among the agents.
The function $\ol{f}_i$ is supposed to represent the  running costs
which can be  dependent on the position size, cash flows, prices of the securities
as well as any relevant news available to each agent. 
The function $\ol{g}_i$ puts some penalty on the position size at the terminal time $T$.
In particular,  we can make the $i$th agent more risk averse
by assigning stronger convexity on $x$ for $\ol{f}_i$ and/or $\ol{g}_i$.

Although the existence of $c^0$ and $c^i$ will not play any significant mathematical role,  
the inclusion of these processes are crucial to construct a meaningful economic model.
\begin{example}
Suppose that the $n$ securities have continuous dividend payments $(c_t^0)_{t\in[0,T)}$ as well as
the rump-sum payment $c_T^0$ at time $T$. In this case,  it may be natural to consider
\be
\begin{split}
\ol{f}_i(t,x,\vp,c^0,c)&=-\langle c^0, x\rangle+\ol{f}_i^\prime (t,x,\vp, c), \\
\ol{g}_i(x,c^0,c)&=-\langle c^0, x\rangle+\ol{g}_i^\prime(x,c), 
\end{split} 
\nn
\ee
with some appropriate measurable functions $\ol{f}_i^\prime$ and $\ol{g}_i^\prime$. 
Here, the first term $\langle c^0,x\rangle$ denotes the benefit from the receipt of the cash flow.
\end{example}

\begin{remark}[ price~($\vp$)-dependence in the cost and coefficient functions ]
Let us emphasize that in the current and previous~\cite{Fujii-Takahashi} works, 
we allow the price-dependence in the running cost $(\ol{f}_i)$ and also in the 
coefficient function $(l_i)$ of the sate $(X^i)$ dynamics. This is particularly because we want to capture self (de)excited behaviors among the agents 
as well as the OTC clients
with respect to the price actions in the market
and to investigate their impacts on the equilibrium price dynamics.
For example, since $\langle \vp_t, X_t^i\rangle $ denotes the mark-to-market value at $t$, 
its higher value is likely to make the $i$th agent happier (i.e. higher utility) and hence implies his/her lower cost.
We will see in Assumptions~\ref{assumption-B} and \ref{assumption-C}
that we need certain conditions among these terms to make the equilibrium well-posed.
Some  economic interpretations are available in the discussion given at the end of Section~\ref{sec-N-equilibrium}.
Another important reason to include $\vp$-dependence is to capture the cash flows from a certain type of securities.
In fact, the foreign currencies and some equities have dividend payments proportional
to their market values.
For example,  if the dividend yield is 
denoted by $\mu$ for this type of asset (say $k$),  we can capture its cash flow by
including  $(-1)\mu \vp_t^k (X_t^i)^k$ in the cost function $\ol{f}_i$.
\end{remark}

In order to discuss the equilibrium among large (or infinite) number of agents,
we need the following concept.
\begin{definition}[price taker]
An agent is called a {\it price taker} if he/she behaves under the assumption 
that there is no price impact from his/her trading. 
\end{definition}
The next assumption is used throughout the current work.
\begin{assumption}
\label{assumption-price-taker}
Every agent $1\leq i\leq N$ behaves as a price taker.
\end{assumption}
This is a very natural assumption if $N$ is sufficiently large since every agent knows that
he/she has only a negligible market share and hence has no capability of affecting the securities price.
In fact, except the special situation\footnote{
For example, if a major financial firm is forced to unwind a huge position within a limited time window, he/she may naturally carry out 
strategic trading by adopting some price impact model.}, this is a common assumption 
used among the practitioners.
For example, if we adopt a certain stochastic price process such as Black-Scholes or SABR model
and then use it for pricing financial derivatives or constructing the optimal portfolio,
then we are actually behaving as price takers.  As one can easily guess,
the vast majority of macro economic problems have also been studied under this assumption.

Note that, under Assumption~\ref{assumption-price-taker},  $(\ref{agent-P})$
becomes the standard optimization,
in which every agent does not consider his/her price impact and just treats $(\vp_t)_{t\in[0,T]}$
as a given price process.
As a result,  it is not difficult to solve it under the  appropriate conditions.
Note however that   the set of  solutions $(\wh{\alpha}^i_t, t\in[0,T], 1\leq i\leq N)$ 
does not satisfy the market clearing condition in general with a given $(\vp_t)_{t\in[0,T]}$.
Our first task is to find an appropriate $(\vp_t)_{t\in[0,T]}$ so that it gives the market clearing equilibrium
among the $N$ price takers.

\subsection{Individual optimization problem}
We are now going to solve the individual optimization problem $(\ref{agent-P})$
for a general given price process for the $n$ securities $(\vp_t)_{t\in[0,T]}\in \mbb{H}^2(\mbb{F};\mbb{R}^n)$.
Let us introduce the following assumptions on the cost functions.
\begin{assumption}
\label{assumption-A}
Uniformly in $1\leq i\leq N$, we assume the following conditions:\\
(i) $\Lambda$ is a positive definite $n\times n$ symmetric matrix with 
$\ul{\lambda}|\theta|^2\leq \langle\theta, \Lambda\theta\rangle\leq \ol{\lambda}|\theta|^2$ for any $\theta\in \mbb{R}^n$
where $0<\ul{\lambda}\leq \ol{\lambda}$ are some constants. \\
(ii) For any $(t,x,\vp,c^0,c)\in [0,T]\times (\mbb{R}^n)^4$, 
\bea
|\ol{f}_i(t,x,\vp,c^0,c)|+|\ol{g}_i(x,c^0,c)|\leq L(1+|x|^2+|\vp|^2+|c^0|^2+|c|^2)~. \nn
\eea
(iii) $\ol{f}_i$ and $\ol{g}_i$ are once continuously differentiable in $x$ and, for any $(t,x,x\pp,\vp,c^0,c)\in [0,T]\times (\mbb{R}^n)^5$,  
\bea
|\part_x\ol{f}_i(t,x\pp,\vp,c^0,c)-\part_x \ol{f}_i(t,x,\vp,c^0,c)|+|\part_x\ol{g}_i(x\pp,c^0,c)-\part_x\ol{g}_i(x,c^0,c)|
\leq L|x\pp-x|~, \nn
\eea
and $|\part_x \ol{f}_i(t,x,\vp,c^0,c)|+|\part_x \ol{g}_i(x,c^0,c)|\leq L(1+|x|+|\vp|+|c^0|+|c|)$.\\
(iv)The functions $\ol{f}_i$ and $\ol{g}_i$ are convex in $x$ in the sense that
\bea
&&\ol{f}_i(t,x\pp,\vp,c^0,c)-\ol{f}_i(t,x,\vp,c^0,c)-\langle x\pp-x, \part_x \ol{f}_i(t,x,\vp,c^0,c)\rangle\geq \frac{\gamma^f}{2}|x\pp-x|^2~, \nn \\
&&\ol{g}_i(x\pp,c^0,c)-\ol{g}_i(x,c^0,c)-\langle x\pp-x,\part_x \ol{g}_i(x,c^0,c)\rangle \geq \frac{\gamma^g}{2}|x\pp-x|^2~, \nn
\eea
for any $(t,x,x\pp,\vp,c^0,c)\in[0,T]\times (\mbb{R}^n)^5$ with some constants $\gamma^f,\gamma^g\geq 0$. \\
(v) For any $(t,\vp,c^0,c)\in[0,T]\times (\mbb{R}^n)^3$,
\bea
|l_i(t,\vp,c^0,c)|+|\sigma_i^0(t,c^0,c)|+|\sigma_i(t,c^0,c)|\leq L(1+|\vp|+|c^0|+|c|). \nn
\eea
(vi) $b \in[0,1)$ is a given constant.
\end{assumption}

\begin{remark}
Note that the condition (iv) in the above assumptions implies 
\bea
&&\bigl\langle x\pp-x, \part_x \ol{f}_i (t,x\pp, \vp,c^0,c)-\part_x \ol{f}_i(t,x,\vp,c^0,c)\bigr\rangle \geq \gamma^f|x\pp-x|^2, \nn \\
&&\bigl\langle  x\pp-x, \part_x \ol{g}_i(x\pp,c^0,c)-\part_x \ol{g}_i(x,c^0,c)\bigr\rangle \geq \gamma^g |x\pp-x|^2,\nn
\eea
which is frequently used in the following analyses.
\end{remark}

The associated (reduced) Hamiltonian~\footnote{Since $\sigma_i^0,\sigma_i$ are independent of the control $\alpha^i$ and also the state $x^i$,
it suffices to use the reduced Hamiltonian for the adjoint equation.} $H_i:[0,T]\times (\mbb{R}^n)^6\rightarrow \mbb{R}$~
\bea
H_i(t,x,y,\alpha,\vp,c^0,c):=\langle y, \alpha+l_i(t,\vp,c^0,c)\rangle+f_i(t,x,\alpha,\vp,c^0,c) \nn
\eea
has a unique minimizer
\bea
\wh{\alpha}(y,\vp):=-\ol{\Lambda}(y+\vp)
\label{eq-alpha}
\eea
where $\ol{\Lambda}:=\Lambda^{-1}$.
The adjoint equation for the $i$th agent arising from the stochastic maximum principle is then given by
\bea
\begin{cases}
dX_t^i=\Bigl(\wh{\alpha}(Y_t^i,\vp_t)+l_i(t,\vp_t, c_t^0,c_t^i)\Bigr)dt+\sigma_i^0(t,c_t^0,c_t^i)dW_t^0+
\sigma_i(t,c_t^0,c_t^i)dW_t^i, \\
dY_t^i=-\part_x\ol{f}_i(t,X_t^i, \vp_t, c_t^0,c_t^i)dt+Z_t^{i,0}dW_t^0+\sum_{j=1}^N Z_t^{i,j}dW_t^j,
\end{cases}
\label{eq-fbsde-single}
\eea
with $X_0^i=\xi^i$ and $Y_T^i:=-b \vp_T+\part_x \ol{g}_i(X_T^i,c_T^0,c_T^i)$.

\begin{theorem}
\label{th-existence-single}
Let Assumptions \ref{assumption-price-taker} and \ref{assumption-A} be in force. 
Then, for given $T>0$ and $(\vp_t)_{t\in[0,T]}\in \mbb{H}^2(\mbb{F};\mbb{R}^n)$, 
the problem $(\ref{agent-P})$ for each agent $1\leq i\leq N$ is uniquely 
characterized by the FBSDE $(\ref{eq-fbsde-single})$ which is strongly solvable 
with a unique solution $(X^i, Y^i, Z^{i,0}, (Z^{i,j})_{j=1}^N)\in \mbb{S}^2(\mbb{F};\mbb{R}^n)\times \mbb{S}^2(\mbb{F};\mbb{R}^n)
\times \mbb{H}^2(\mbb{F};\mbb{R}^{n\times d_0})\times (\mbb{H}^2(\mbb{F};\mbb{R}^{n\times d}))^N$. 
\begin{proof}
This is essentially the same as Theorem~3.1 in \cite{Fujii-Takahashi}.
Since the cost functions are jointly convex with $(x,\a)$ and strictly convex in $\a$, the problem 
is the special situation investigated in Section 1.4.4 in \cite{Carmona-Delarue-2}. 
The existence of a unique solution to the FBSDE can be 
proved in a similar way as Theorem 1.60 in the same reference.
Because of the differentiability as well as the convexity of cost functions, both the necessary and the sufficient conditions
of the Pontryagin's maximum principle are satisfied. Hence, together with the unique existence of the solution to the FBSDE,
the optimal solution is uniquely characterized in terms of the FBSDE.\footnote{The existence of the 
solution to the FBSDE can also be proved via Peng-Wu's method \cite{Peng-Wu}.}
\end{proof}
\end{theorem}

\subsection{Market clearing equilibrium among $N$ agents}
From Theorem~\ref{th-existence-single}, the optimal trading strategy of the agent $i$ for a given $(\vp_t)_{t\in[0,T]}$ is 
\bea
\wh{\alpha}_t^i:=-\ol{\Lambda}(Y_t^i+\vp_t), \quad t\in[0,T].\nn
\eea
Since the market clearing requires $\sum_{i=1}^N \wh{\alpha}_t^i=0$, the market price 
needs to satisfy
\bea
\vp_t=-\frac{1}{N}\sum_{i=1}^N Y_t^i=-\mg{m}\bigl((Y_t^i)\bigr), \quad t\in[0,T].
\label{eq-market-price}
\eea
The above expression implies that every agent interacts in a symmetric way through the market price.
This observation motivates us to consider 
the following $N$ coupled system of FBSDEs, which is obtained by substituting the price process $(\vp_t)_{t\in[0,T]}$ in 
$(\ref{eq-fbsde-single})$ for every $1\leq i\leq N$ with the one given in $(\ref{eq-market-price})$.
\bea
\begin{cases}
1\leq i \leq N,  \\
dX_t^i:=\Bigl\{ \wh{\alpha}\bigl(Y^i_t, -\mg{m}\bigl((Y_t^j)\bigr)\bigr)
+l_i\bigl(t, -\mg{m}\bigl((Y_t^j)\bigr), c_t^0,c_t^i\bigr)\Bigr\}dt+\sigma_i^0(t,c_t^0, c_t^i)dW_t^0+\sigma_i(t,c_t^0,c_t^i)dW_t^i, \\
dY_t^i=-\part_x \ol{f}_i\Bigl(t,X_t^i, -\mg{m}\bigl((Y_t^j)\bigr), c_t^0,c_t^i\Bigr)dt+
Z_t^{i,0}dW_t^0+\sum_{j=1}^N Z_t^{i,j}dW_t^j,   \\
\end{cases}
\label{eq-fbsde-N}
\eea
$t\in[0,T]$ with
\be
X_0^i=\xi^i,  \quad Y_T^i=\frac{b}{1-b}\mg{m}\bigl((\part_x \ol{g}_j(X_T^j,c_T^0,c_T^j))\bigr)+\part_x \ol{g}_i(X_T^i,c_T^0,c_T^i).
\nn
\ee
Let us mention about the terminal condition.
Since we have $(\ref{eq-market-price})$, $Y_T^i$ must satisfy
\bea
Y_T^i=b \frac{1}{N}\sum_{j=1}^N Y_T^j+\part_x \ol{g}_i(X_T^i,c_T^0,c_T^i).\nn
\eea
Summing over $1\leq i\leq N$, we can solve  $\frac{1}{N}\sum_{j=1}^N Y_T^j$ as $
\mg{m}\bigl((Y_T^j))=\frac{1}{1-b}\mg{m}\bigl((\part_x \ol{g}_j(X_T^j,c_T^0,c_T^j))\bigr)$.
Substituting the result into the above terminal condition, we get the desired result.
The next theorem reveals the crucial importance of the above system of FBSDEs.
\begin{theorem}
\label{th-N-equilibrium}
Let Assumptions \ref{assumption-price-taker} and \ref{assumption-A} be in force.
The market clearing equilibrium among the $N$ agents with a square integrable price process 
$(\vp_t)_{t\in[0,T]}\in \mbb{H}^2(\mbb{F};\mbb{R}^n)$  exists if and only if there exists a solution 
 $(X^i, Y^i, Z^{i,0}, (Z^{i,j})_{j=1}^N)\in \mbb{S}^2(\mbb{F};\mbb{R}^n)\times \mbb{S}^2(\mbb{F};\mbb{R}^n)
\times \mbb{H}^2(\mbb{F};\mbb{R}^{n\times d_0})\times (\mbb{H}^2(\mbb{F};\mbb{R}^{n\times d}))^N$, $1\leq i\leq N$ 
to the $N$-coupled system of FBSDEs
$(\ref{eq-fbsde-N})$. 
\begin{proof}
Suppose that there exists a market clearing equilibrium among the $N$ agents with a square integrable price process $(\vp_t)_{t\in[0,T]}$.
Then, from Theorem~\ref{th-existence-single},  the solution $\bigl(X^i, Y^i, Z^{i,0}, (Z^{i,j})_{j=1}^N\bigr)$ to $(\ref{eq-fbsde-single})$ for 
each agent $1\leq i\leq N$ satisfies the equality $(\ref{eq-market-price})$. Hence, we see the system of FBSDEs $(\ref{eq-fbsde-N})$
is in fact solved by the same set of square integrable processes $\bigl(X^i, Y^i, Z^{i,0}, (Z^{i,j})_{j=1}^N\bigr)$, $1\leq i\leq N$.

Conversely, suppose that the $N$-coupled system of FBSDEs $(\ref{eq-fbsde-N})$ 
has a square integrable solution  $\bigl(X^i, Y^i, Z^{i,0}, (Z^{i,j})_{j=1}^N\bigr)$, $1\leq i\leq N$. 
Let us define the price process $(\vp_t)_{t\in[0,T]}$ by 
$(\ref{eq-market-price})$.
For each $1\leq i\leq N$, let us denote  by $\bigl(x^i, y^i, z^{i,0}, (z^{i,j})_{j=1}^N\bigr)$ 
the solution to $(\ref{eq-fbsde-single})$ of the individual agent problem with this price process $(\vp_t)_{t\in[0,T]}$ 
as an input. Since the solution is unique by Theorem~\ref{th-existence-single}, 
we see that  $y^i=Y^i$ in $\mbb{S}^2(\mbb{F};\mbb{R}^n)$ for every $1\leq i\leq N$.
As a result, $\vp_t=-\mg{m}\bigl((y_t^i)\bigr)$ and hence the market clearing condition
\be
\sum_{i=1}^N \wh{\alpha}(y_t^i, \vp_t)=0, \quad t\in[0,T] \nn
\ee
is in fact satisfied. 
\end{proof}
\end{theorem}

Notice that the linear-quadratic dependence of $\alpha$ in $(\ref{eq-cost-functions})$ plays 
a crucial role to obtain a simple expression $(\ref{eq-market-price})$.  In theory, we may allow
more general $\alpha$ dependence in the cost function $f$ in particular 
by adopting the conditions used in \cite[Lemma 3.3]{Carmona-Delarue-1} which guarantees
the minimizer $\wh{\alpha}$ of the Hamiltonian is Lipschitz-continuous with respect to $(x,y)$.
However, it will still make the treatment of the market-clearing condition 
more complicated and technical. In this work, we will not follow this line of arguments and treat, if necessary,  the linear-quadratic 
form as a useful approximation for the analysis.  See  the method of calculus of variations  in the recent publication~\cite{Gomes-etal-2} as
a different approach for this issue.

We now introduce a new set of assumptions to prove the existence of the solution to $(\ref{eq-fbsde-N})$.
\begin{assumption}
\label{assumption-B}
(i) For any $(t,x,\vp,\vp\pp, c^0,c)\in[0,T]\times (\mbb{R}^n)^5$, 
\be
|l_i(t,\vp,c^0,c)-l_i(t,\vp\pp,c^0,c)|\leq L |\vp-\vp\pp|, \nn
\ee
and moreover,  there exists some nonnegative constant $L_{\vp}$ such that
\be
|\part_x \ol{f}_i(t,x,\vp,c^0,c)-\part_x\ol{f}_i(t,x,\vp\pp,c^0,c)|\leq L_\vp|\vp-\vp\pp|, \nn
\ee
for every $1\leq i\leq N$.\\
(ii) For any $(t,c^0)\in[0,T]\times \mbb{R}^n$ and $(x^i,x^{i\prime},c^i)\in (\mbb{R}^n)^3, 1\leq i\leq N$, 
the functions $(l_i)_{i=1}^N$ satisfy with some $\gamma^l>0$
\bea
\frac{1}{N}\sum_{i=1}^N \bigl\langle l_i\bigl(t,\mg{m}((x^j)), c^0,c^i\bigr)
-l_i\bigl(t,\mg{m}((x^{j\prime})),c^0,c^i\bigr), x^i-x^{i\prime}\bigr\rangle
\geq \gamma^l \bold{1}_{\{L_\vp>0\}} \bigl|\mg{m}\bigl((x^i-x^{i\prime})\bigr)\bigr|^2. \nn
\eea
(iii) There exists a strictly positive constant $\gamma$ satisfying
\bea
0<\gamma\leq \Bigl(\gamma^f-\frac{L_\vp^2}{4\gamma^l}\Bigr)\wedge \gamma^g~. \nn
\eea
Moreover, the functions $(\ol{g}_i)_{i=1}^N$ satisfy for any $c^0\in\mbb{R}^n$
and  $(x^i,x^{i\prime},c^i)\in (\mbb{R}^n)^3, 1\leq i\leq N$, 
\bea
\frac{b}{1-b}\sum_{i=1}^N
\bigl\langle \mg{m}\bigl((\part_x\ol{g}_j(x^j,c^0,c^j))\bigr)-\mg{m}\bigl((\part_x \ol{g}_j(x^{j\prime},c^0,c^j))\bigr),
x^i-x^{i\prime}\bigr\rangle\geq (\gamma-\gamma^g) \sum_{i=1}^N |x^i-x^{i\prime}|^2. \nn
\eea
\end{assumption}

\begin{remark}
\label{remark-monotonicity}
In economic terms, the monotone condition (ii) can be interpreted in a very natural way.
It basically tells that the demand for the securities from the OTC clients of the agents decreases when the 
market price rises. Let us provide the simplest example of the functions $(l_i)_i$ that satisfy (ii); assume that $l_i$ has a separable form 
$l_i(t,x,c^0,c^i)=h(t,x)+h_i(t,c^0,c^i)$ and also that the common function $h$ is strictly monotone in $x$.
Then, one can easily check that (ii) is satisfied.  

Combined with Assumption~\ref{assumption-A} (iv), the above condition $(iii)$ implies the $\gamma$-convexity  
(see $(\ref{ineq-G})$ below) with respective to the function:
\be
\frac{b}{1-b}\mg{m}\bigl((\ol{g}_j(x^j,c^0,c^j)\bigr)+\ol{g}_i(x^i,c^0,c^j), \quad 1\leq i\leq N, \nn
\ee
with $\gamma\leq \gamma^g$. 
As is well-known, requiring the convexity in the terminal function is standard for optimization problems. In particular,
it is used to guarantee the sufficiency of the Pontryagin's maximum principle as well as the well-posedness of the associated FBSDE.
The second inequality in Assumption~\ref{assumption-A} (iv) serves this role for the individual problem.
However, this convexity can possibly be destroyed if we  include new interaction terms induced by the market clearing condition.
The inequality in Assumption~\ref{assumption-B} (iii)  prevents it from happening.
The condition can be satisfied, for example,  if  $(\part_x \ol{g}_i)_i$ have a similar separable structure as explained for $(l_i)_{i}$
in the last paragraph.
\end{remark}

For notational convenience for later analyses, let us introduce the following functions:
$B_i:[0,T]\times \mbb{R}^n\times \calp(\mbb{R}^n)\times (\mbb{R}^n)^2\rightarrow \mbb{R}^n$,
$F_i:[0,T]\times \mbb{R}^n \times \calp(\mbb{R}^n)\times (\mbb{R}^n)^2\rightarrow \mbb{R}^n$
and $G_i:\calp(\mbb{R}^n)\times (\mbb{R}^n)^3\rightarrow \mbb{R}^n$, $1\leq i\leq N$ by
\be
\begin{split}
&B_i(t,x,\mu,c^0,c):=\wh{\alpha}(x,-m(\mu))+l_i(t,-m(\mu),c^0,c),  \\
&F_i(t,x,\mu,c^0,c):=-\part_x \ol{f}_i(t,x,-m(\mu),c^0,c),  \\
&G_i(\mu, x,c^0,c):=\frac{b}{1-b}m(\mu)+\part_x \ol{g}_i(x,c^0,c), 
\end{split}
\label{def-BFG}
\ee
for any $(t,x,\mu,c^0,c)\in[0,T]\times \mbb{R}^n\times \calp(\mbb{R}^n)\times (\mbb{R}^n)^2$.

\begin{theorem}
\label{th-fbsde-N-existence}
Let Assumptions~\ref{assumption-A} and \ref{assumption-B} be in force.
Then, for any $T>0$,  the $N$-coupled system of FBSDEs $(\ref{eq-fbsde-N})$ 
has a unique strong solution $(X^i,Y^i,Z^{i,0},(Z^{i,j})_{j=1}^N) \in \mbb{S}^2(\mbb{F};\mbb{R}^n)\times \mbb{S}^2(\mbb{F};\mbb{R}^n)
\times \mbb{H}^2(\mbb{F};\mbb{R}^{n\times d_0})\times (\mbb{H}^2(\mbb{F};\mbb{R}^{n\times d}))^N$, $1\leq i\leq N$.
\begin{proof}
We can prove the claim by a simple modification of Theorem 6.2 in \cite{Fujii-Takahashi}.
We make the following hypothesis: {\it there exists some constant $\vr\in[0,1)$ such that
for any $I^{b,i}, I^{f,i}\in \mbb{H}^2(\mbb{F};\mbb{R}^n)$
and for any $\eta^i\in \mbb{L}^2(\calf_T;\mbb{R}^n)$, there exists a unique strong solution 
$(x^{\vr,i}, y^{\vr,i}, z^{\vr,i,0}, (z^{\vr,i,j})_{j=1}^N)\in \mbb{S}^2\times \mbb{S}^2\times \mbb{H}^2
\times (\mbb{H}^2)^N$, $1\leq i\leq N$ to the $N$-coupled system of FBSDEs:
\bea
\begin{cases}
dx_t^{\vr,i}=\bigl(\vr B_i(t,y_t^{\vr,i}, \mu^{\vr,N}_t,c_t^0,c_t^i)+I_t^{b,i}\bigr)dt+
\sigma_i^0(t,c_t^0,c_t^i)dW_t^0+\sigma_i(t,c_t^0,c_t^i)dW_t^i,  \\
dy_t^{\vr,i}=-\bigl((1-\vr)\gamma x_t^{\vr,i}-\vr F_i(t,x_t^{\vr,i},\mu_t^{\vr,N},c_t^0,c_t^i)+I_t^{f,i}\bigr)dt+
z_t^{\vr,i,0}dW_t^0+\sum_{j=1}^N z_t^{\vr,i,j}dW_t^j, 
\end{cases} \nn
\eea
for $t\in[0,T]$ with $x_0^{\vr,i}=\xi^i$ and $y_T^{\vr,i}=\vr G_i(\mu^{\vr,N}_g,x_T^{\vr,i},c_T^0,c_T^i)+(1-\vr)x_T^{\vr,i}+\eta^i$.
Here,
\bea
\mu_t^{\vr,N}:=\frac{1}{N}\sum_{i=1}^N \del_{y_t^{\vr,i}},\quad \mu_g^{\vr,N}:=\frac{1}{N}\sum_{i=1}^N \del_{\part_x \ol{g}_i(x_T^{\vr,i},c_T^i,c_T^i)} \nn
\eea
denote the empirical measures.} 

Notice that the system reduces to the $N$ decoupled FBSDEs
when $\vr=0$. Hence, the hypothesis trivially holds for $\vr=0$.
Now, for some constant $\zeta\in(0,1)$, we define a map 
\bea
&&\bigl(\mbb{S}^2\times \mbb{S}^2\times \mbb{H}^2\times(\mbb{H}^2)^N\bigr)^N\ni 
\bigl(x^i,y^i,z^{i,0},(z^{i,j})_{j=1}^N\bigr)_{i=1}^N \nn \\
&&\qquad\qquad \mapsto \bigl(X^i,Y^i,Z^{i,0},(Z^{i,j})_{j=1}^N\bigr)_{i=1}^N \in \bigl(\mbb{S}^2\times \mbb{S}^2\times \mbb{H}^2\times(\mbb{H}^2)^N\bigr)^N 
\label{map-continuation}
\eea
by 
\bea
\begin{cases}
dX_t^i=\bigl[\vr B_i(t,Y_t^i, \mu_t^N,c_t^0,c_t^i)+\zeta B_i(t,y_t^i, \nu_t^{N},c_t^0,c_t^i)+I_t^{b,i}\bigr]dt 
+\sigma_i^0(t,c_t^0,c_t^i)dW_t^0+\sigma_i(t,c_t^0,c_t^i)dW_t^i, \\
dY_t^i=-\Bigl[(1-\vr)\gamma X_t^i-\vr F_i(t,X_t^i,\mu_t^N,c_t^0,c_t^i)+
\zeta\bigl(-\gamma x_t^i-F_i(t,x_t^i,\nu_t^N,c_t^0,c_t^i)\bigr)+I_t^{f,i}\Bigr]dt \\
\hspace{10mm}+Z_t^{i,0}dW_t^0+\sum_{j=1}^N Z_t^{i,j}dW_t^j, 
\end{cases} \nn
\eea
with $X_0^i=\xi$ and $Y_T^i=\vr G_i(\mu^N_g,X_T^i,c_T^0,c_T^i)+(1-\vr)X_T^i+\zeta\bigl(
G_i(\nu^N_g, x_T^i,c_T^0,c_T^i)-x_T^i\bigr)+\eta^i$.
Here, the measure arguments are defined by
\bea
&&\mu_t^N:=\frac{1}{N}\sum_{i=1}^N \del_{Y_t^i}, \qquad \nu_t^N:=\frac{1}{N}\sum_{i=1}^N \del_{y_t^i}, \nn \\
&&\mu_g^N:=\frac{1}{N}\sum_{i=1}^N \del_{\part_x \ol{g}_i(X_T^i,c_T^0,c_T^i)}, \quad
\nu_g^N:=\frac{1}{N}\sum_{i=1}^N \del_{\part_x \ol{g}_i(x_T^i,c_T^0,c_T^i)}. \nn
\eea
Thanks for the hypothesis, there exists a unique solution $\bigl(X^i,Y^i,Z^{i,0},(Z^{i,j})_{j=1}^N\bigr)_{i=1}^N$
and hence the map $(\ref{map-continuation})$ is well-defined.

Consider the two set of inputs $\bigl(x^i,y^i,z^{i,0},(z^{i,j})_{j=1}^N\bigr)_{i=1}^N$
and $\bigl(x^{i\prime},y^{i\prime},z^{i,0\prime}, (z^{i,j\prime})_{j=1}^N\bigr)_{i=1}^N$, and 
then denote the corresponding solution to the previous FBSDEs by $\bigl(X^i,Y^i,Z^{i,0},(Z^{i,j})_{j=1}^N\bigr)_{i=1}^N$
and $\bigl(X^{i\prime},Y^{i\prime},Z^{i,0\prime},(Z^{i,j\prime})_{j=1}^N\bigr)_{i=1}^N$, respectively.
Put $\Del X^i:=X^i-X^{i\prime}, \Del Y^i:=Y^i-Y^{i\prime}$, etc.
We have the following monotone conditions:
\be
\begin{split}
&\sum_{i=1}^N \bigl\langle B_i(t, Y_t^i, \mu_t^N, c_t^0, c_t^i)-B_i(t, Y_t^{i\prime}, \mu_t^{\prime N},c_t^0,c_t^i), \Del Y_t^i\bigr\rangle \\
&=-\sum_{i=1}^N \langle \ol{\Lambda}\Del Y_t^i, \Del Y_t^i\rangle+N\langle \ol{\Lambda}\mg{m}((\Del Y_t^i)), \mg{m}((\Del Y_t^i))\rangle \\
&\quad +\sum_{i=1}^N\langle l_i(t,-\mg{m}((Y_t^j)),c_t^0,c_t^i)-l_i(t,-\mg{m}((Y_t^{j\prime})),c_t^0,c_t^i), \Del Y_t^i\rangle \\
& \leq -N\gamma^l \bold{1}_{\{L_\vp>0\}}\bigl|\mg{m}((\Del Y_t^i))\bigr|^2,
\end{split}
\label{ineq-B}
\ee
where Cauchy-Schwarz  inequality and Assumption~\ref{assumption-B}(ii) were used. Similarly, Assumption~\ref{assumption-A}(iv)
and Assumption~\ref{assumption-B}(i) imply
\be
\begin{split}
&\sum_{i=1}^N \bigl\langle F_i(t,X_t^i, \mu_t^N,c_t^0,c_t^i)
-F_i(t,X_t^{i\prime}, \mu_t^{\prime N}, c_t^0,c_t^i),\Del X_t^i\bigr\rangle  \\
&=-\sum_{i=1}^N\bigl\langle  \part_x \ol{f}_i(t, X_t^i,-\mg{m}((Y_t^j)), c^0_t,c_t^i)
-\part_x\ol{f}_i(t,X_t^{i\prime}, -\mg{m}((Y_t^j)), c^0_t,c_t^i),\Del X_t^i\bigr\rangle \\
&\quad-\sum_{i=1}^N\bigl\langle \part_x\ol{f}_i(t,X_t^{i\prime}, -\mg{m}((Y_t^j)), c^0_t,c_t^i)
-\part_x\ol{f}_i(t,X_t^{i\prime}, -\mg{m}((Y_t^{j\prime})), c^0_t,c_t^i),\Del X_t^i\bigr\rangle  \\
&\leq -\gamma^f \sum_{i=1}^N |\Del X_t^i|^2+\sum_{i=1}^N L_\vp|\mg{m}((\Del Y_t^j))||\Del X_t^i|  \\
&\leq -\Bigl(\gamma^f-\frac{L_\vp^2}{4\gamma^l}\Bigr)\sum_{i=1}^N |\Del X_t^i|^2+N\gamma^l \bold{1}_{\{L_\vp>0\}}|\mg{m}((\Del Y_t^i))|^2.
\end{split}
\label{ineq-F}
\ee
Assumption~\ref{assumption-A}(iv) and Assumption~\ref{assumption-B}(iii) immediately yield
\bea
&&\sum_{i=1}^N \bigl\langle G_i(\mu_g^N, X_T^i, c_T^0,c_T^i)-G_i(\mu_g^{\prime N}, X_T^{i\prime},c_T^0,c_T^i),\Del X_T^i\bigr\rangle
\geq \gamma \sum_{i=1}^N |\Del X_T^i|^2.
\label{ineq-G}
\eea

Since $\Del X_0^i=0$, a simple application of \Ito-formula to $\sum_{i=1}^N \langle \Del X^i, \Del Y^i\rangle $ yields
\be
\begin{split}
&\sum_{i=1}^N \mbb{E}\bigl[\langle \Del X_T^i, \Del Y_T^i\rangle\bigr]=
-(1-\vr)\gamma \mbb{E}\int_0^T \sum_{i=1}^N |\Del X_t^i|^2 dt \\
&\qquad+\vr\mbb{E}\int_0^T \sum_{i=1}^N \bigl\langle B_i(t,Y_t^i, \mu_t^N, c_t^0,c_t^i)-
B_i(t,Y_t^{i\prime}, \mu_t^{\prime N}, c_t^0,c_t^i),\Del Y_t^i\bigr\rangle dt  \\
&\qquad+\vr\mbb{E}\int_0^T \sum_{i=1}^N \bigl\langle F_i(t,X_t^i, \mu_t^N,c_t^0,c_t^i)
-F_i(t,X_t^{i\prime}, \mu_t^{\prime N}, c_t^0,c_t^i),\Del X_t^i\bigr\rangle dt \\
&\qquad+\zeta\mbb{E}\int_0^T \sum_{i=1}^N \bigl\langle B_i(t,y_t^i, \nu_t^N c_t^0,c_t^i)
-B_i(t,y_t^{i\prime},\nu_t^{\prime N},c_t^0,c_t^i),\Del Y_t^i\bigr\rangle dt  \\
&\qquad+\zeta \mbb{E}\int_0^T
\sum_{i=1}^N \bigl\langle \gamma\Del x_t^i+F_i(t,x_t^i,\nu_t^N,c_t^0,c_t^i)
-F_i(t,x_t^{i\prime},\nu_t^{\prime N}, c_t^0,c_t^i),\Del X_t^i\bigr\rangle dt~. 
\end{split}\nn
\ee
Using the inequalities $(\ref{ineq-B})$ and $(\ref{ineq-F})$, the Lipschitz continuity for $(B,F)$,
and Assumption~\ref{assumption-B}(iii), we obtain,  with some constant $C$ independent of  $(\vr, N)$, that
\be
\begin{split}
&\sum_{i=1}^N \mbb{E}\bigl[\langle \Del X_T^i, \Del Y_T^i\rangle\bigr] \leq -\gamma \mbb{E}\int_0^T \sum_{i=1}^N |\Del X_t^i|^2 dt  \\
&\hspace{10mm} +\zeta C\mbb{E}\int_0^T \sum_{i=1}^N \Bigl[\bigl(|\Del y_t^i|+|\mg{m}((\Del y_t^j))|\bigr)|\Del Y_t^i|
+\bigl(|\Del x_t^i|+|\mg{m}((\Del y_t^j))|\bigr)|\Del X_t^i|\Bigr]dt. 
\end{split}
\nn
\ee
Using $(\ref{ineq-G})$ and Assumption~\ref{assumption-A}(iii), we get 
\be
\begin{split}
&\sum_{i=1}^N \mbb{E}\bigl[\langle \Del X_T^i, \Del Y_T^i\rangle\bigr] =\vr\mbb{E}\sum_{i=1}^N \bigl\langle G_i(\mu_g^N, X_T^i,c_T^0,c_T^i)-G_i(\mu_g^{\prime N}, X_T^{i\prime}, c_T^0,c_T^i),\Del X_T^i\bigr\rangle \\
&\quad+(1-\vr)\mbb{E}\sum_{i=1}^N \bigl\langle \Del X_T^i, \Del X_T^i\bigr\rangle
+\zeta\mbb{E}\sum_{i=1}^N \bigl\langle G_i(\nu_g^N,x_T^i,c_T^0,c_T^i)-G_i(\nu_g^{\prime N}, x_T^{i\prime},c_T^0,c_T^i)-\Del x_T^i,
\Del X_T^i\bigr\rangle  \\
&\quad \geq
(\vr \gamma+(1-\vr))\mbb{E}\Bigl[\sum_{i=1}^N |\Del X_T^i|^2\Bigr]-\zeta C\mbb{E}\Bigl[\sum_{i=1}^N 
\bigl(|\Del x_T^i|+\mg{m}((|\Del x_T^j|))\bigr)|\Del X_T^i|\Bigr].
\end{split}\nn
\ee
Here,  we have used a simple fact that 
\be
\bigl|\mg{m}\bigl((\part_x \ol{g}_i(x_T^i,c_t^0,c_T^i))\bigr)-\mg{m}\bigl((\part_x \ol{g}_i(x_T^{i\prime},c_T^0,c_T^i))\bigr)\bigr|
\leq L\mg{m}((|\Del x_T^i|)).\nn
\ee

With $\gamma_c:=\min(1,\gamma)$, we have $0<\gamma_c\leq \vr\gamma+(1-\vr)$, and hence the above two estimates give
\be
\begin{split}
&\gamma_c \sum_{i=1}^N \mbb{E}\Bigl[|\Del X_T^i|^2+\int_0^T |\Del X_t^i|^2 dt\Bigr]
\leq \zeta C\sum_{i=1}^N \mbb{E}\Bigl[\bigl(|\Del x_T^i|+\mg{m}((|\Del x_T^j|))\bigr)|\Del X_T^i|\Bigr] \\
&\qquad+\zeta C\mbb{E}\int_0^T \sum_{i=1}^N \Bigl[(|\Del y_t^i|+|\mg{m}((\Del y_t^j))|)|\Del Y_t^i|+
\bigl(|\Del x_t^i|+|\mg{m}((\Del y_t^j))|\bigr)|\Del X_t^i|\Bigr]dt.  
\end{split}\nn
\ee
Since $|\mg{m}((x^i))|^2\leq \frac{1}{N}\sum_{i=1}^N |x^i|^2$,  we obtain from  Young's inequality, 
\be
\sum_{i=1}^N \mbb{E}\Bigl[|\Del X_T^i|^2+\int_0^T |\Del X_t^i|^2 dt\Bigr]
\leq \zeta C\sum_{i=1}^N \mbb{E}\Bigl[|\Del x_T^i|^2+\int_0^T \bigl(|\Del x_t^i|^2+|\Del y_t^i|^2+|\Del Y_t^i|^2\bigr)dt\Bigr].
\label{eq-X-estimate}
\ee

Let us now treat $(X^i, X^{i\prime})_{i=1}^N$ as the exogenous inputs. Then the standard stability result
for the Lipschitz BSDEs (see, for example, Theorem~4.2.3 in \cite{Zhang-BSDE}) implies
\be
\begin{split}
&\sum_{i=1}^N \mbb{E}\Bigl[\sup_{t\in[0,T]}|\Del Y_t^i|^2+\int_0^T |\Del Z_t^{i,0}|^2 dt+\sum_{j=1}^N\int_0^T |\Del Z_t^{i,j}|^2dt\Bigr]  \\
&\qquad \leq C\sum_{i=1}^N \mbb{E}\Bigl[|\Del X_T^i|^2+\int_0^T |\Del X_t^i|^2 dt\Bigr]+
\zeta C\sum_{i=1}^N \mbb{E}\Bigl[|\Del x_T^i|^2+\int_0^T \bigl(|\Del x_t^i|^2+|\Del y_t^i|^2\bigr)dt\Bigr]. 
\end{split}\nn
\ee
Using $(\ref{eq-X-estimate})$ and small $\zeta$, we obtain
\be
\begin{split}
&\sum_{i=1}^N \mbb{E}\Bigl[\sup_{t\in[0,T]}|\Del Y_t^i|^2+\int_0^T |\Del Z_t^{i,0}|^2 dt+\sum_{j=1}^N\int_0^T |\Del Z_t^{i,j}|^2dt\Bigr]  \\
&\qquad \leq \zeta C \sum_{i=1}^N \mbb{E}\Bigl[|\Del x_T^i|^2+\int_0^t \bigl(|\Del x_t^i|^2+|\Del y_t^i|^2\bigr)dt\Bigr]. 
\end{split}
\label{eq-Y-Z-estimate}
\ee
Similarly, by treating $(Y^i,Y^{i\prime})_{i=1}^N$ as the exogenous inputs, the standard stability result for the
Lipschitz SDEs gives
\be
\sum_{i=1}^N \mbb{E}\Bigl[\sup_{t\in[0,T]}|\Del X_T^i|\Bigr]
\leq \zeta C\sum_{i=1}^N \mbb{E}\int_0^T |\Del y_t^i|^2 dt+C\sum_{i=1}^N \mbb{E}\int_0^T |\Del Y_t^i|^2 dt. 
\label{eq-X-es2}
\ee
Therefore, from $(\ref{eq-Y-Z-estimate})$ and $(\ref{eq-X-es2})$, we obtain
\be
\begin{split}
&\sum_{i=1}^N \mbb{E}\Bigl[\sup_{t\in[0,T]}|\Del X_t^i|^2+\sup_{t\in[0,T]}|\Del Y_t^i|^2+\int_0^T |\Del Z_t^{i,0}|^2 dt+\sum_{j=1}^N\int_0^T |\Del Z_t^{i,j}|^2dt\Bigr]  \\
&\qquad \leq \zeta C \sum_{i=1}^N \mbb{E}\Bigl[\sup_{t\in[0,T]}|\Del x_t^i|^2+\sup_{t\in[0,T]}|\Del y_t^i|^2\Bigr]. 
\end{split}\nn
\ee
Thus for small $\zeta>0$, which can be taken independently from $\vr$, the map $(\ref{map-continuation})$
becomes a strict contraction. Hence the Banach fixed point theorem implies that the initial hypothesis holds for $(\vr+\zeta)$.
Repeating the procedures, we see the hypothesis holds with $\vr=1$. This establishes the existence of a solution.
The uniqueness is a direct result of the Banach's fixed point  theorem. 
\end{proof}
\end{theorem}

To the best of the authors' knowledge, the current work is the first example which directly applies the method~\cite{Peng-Wu}
to prove the existence of market clearing equilibrium in an incomplete market with finite number of agents.
Although the  mathematical technique itself is the standard one, this is an interesting result in itself.
This is particularly because that the existing literature studying finite population markets
always adopts the exponential-type utility function with respect to the terminal wealth\footnote{See, for examples, \cite{Zitkovic-1, Xing-Zitkovic,
Weston-Zitkovic}.}, and also because that the heterogeneity among the agents is modeled solely by the
risk-tolerance coefficient of the utility function\footnote{
See recent publications~\cite{Gomes-etal-1, Gomes-etal-2} as interesting exceptions.}.
In contrast, thanks to the generality of \cite{Peng-Wu}, we can suppose that each agent has a different cost (or utility) function 
(not only in the coefficients but also in its functional form)
as long as the appropriate convexity and monotonicity conditions are satisfied uniformly among the agents.

For later use, we give the stability result for the FBSDEs.
\begin{proposition}
\label{prop-N-stability}
Given two set of inputs $(\xi^i,c^0,c^i)_{i=1}^N$, $(\xi^{i\prime},c^{0\prime}, c^{i\prime})_{i=1}^N$,
and the coefficients functions $(l_i,\sigma_i^0,\sigma_i, f_i, g_i)_{i=1}^N$, $(l_i^\prime, \sigma_i^{0\prime}, \sigma_i^\prime, f_i\pp,g_i\pp)$
satisfying Assumptions~\ref{assumption-A} and \ref{assumption-B}, let us denote the corresponding solutions
to $(\ref{eq-fbsde-N})$  by $(X^i, Y^i,Z^{i,0},(Z^{i,j})_{j=1}^N)_{i=1}^N$
and $(X^{i\prime},Y^{i\prime}, Z^{i,0\prime}, (Z^{i,j\prime})_{j=1}^N)_{i=1}^N$, respectively.
Then, for $\Del X^i:=X^i-X^{i\prime}, \Del Y^i:=Y^i-Y^{i\prime}, \Del Z^{i,j}:=Z^{i,j}-Z^{i,j\prime}$, $1\leq i, j\leq N$, 
we have
\be
\begin{split}
&\sum_{i=1}^N \mbb{E}\Bigl[\sup_{t\in[0,T]}|\Del X_t^i|^2+\sup_{t\in[0,T]}|\Del Y_t^i|^2+
\int_0^T \bigl(|\Del Z_t^{i,0}|^2+\sum_{j=1}^N|\Del Z_t^{i,j}|^2\bigr)dt\Bigr] \\
&\qquad \leq C\sum_{i=1}^N \mbb{E}\Bigl[|\Del \xi^i|^2+|\del G_i|^2+
\int_0^T \bigl(|\del F_i(t)|^2+|\del B_i(t)|^2+|\del \sigma_i^0(t)|^2+|\del \sigma_i(t)|^2\bigr)dt\Bigr], 
\end{split}\nn
\ee
with some constant $C$ depending only on the Lipschitz constants, $\del, \ul{\lambda}$ and $\gamma$. Here, 
\be
\begin{split}
&\del B_i(t):=B_i(t,Y_t^{i\prime},\mu_t^{\prime N}, c_t^0,c_t^i)-B_i^\prime(t,Y_t^{i\prime}, \mu_t^{\prime N}, c_t^{0\prime},c_t^{i\prime}),  \\
&\del F_i(t):=F_i(t,X_t^{i\prime}, \mu_t^{\prime N},c_t^0,c_t^i)-F_i^\prime(t,X_t^{i\prime},\mu_t^{\prime N}, c_t^{0\prime},c_t^{i\prime}),  \\
&\del G_i:=G_i(\mu_g^{\prime N}, X_T^{i\prime},c_T^0,c_T^i)-G_i^\prime(\mu_{g\prime}^{\prime N}, X_T^{i\prime}, c_T^{0\prime}, c_T^{i\prime}), \\
&(\del \sigma_i^0,\del \sigma_i)(t):=\bigl((\sigma_i^0, \sigma_i)(t,c_t^0,c_t^i)-
(\sigma_i^{0\prime}, \sigma_i^\prime)(t,c_t^{0\prime}, c_t^{i\prime})\bigr),
\end{split}\nn
\ee
for $t\in[0,T]$ and $1\leq i\leq N$. Here $B_i^\prime,F_i^\prime$ and $G_i^\prime$ are defined as $(\ref{def-BFG})$ with primed variables.
The measure arguments are defined by $\mu_t^{\prime N}:=\frac{1}{N}\sum_{i=1}^N \del_{Y_t^{i\prime}}$,  $\mu_g^{\prime N}:=\frac{1}{N}\sum_{i=1}^N \del_{\part_x \ol{g}_i}(X_T^{i\prime},c_T^0,c_T^i)$ and $\mu_{g\prime}^{\prime N}:=\frac{1}{N}\sum_{i=1}^N \del_{\part_x \ol{g}_i^\prime}(X_T^{i\prime},c_T^{0\prime},c_T^{i\prime})$.
\begin{proof}
One can prove the claim exactly in the same way as \cite[Proposition 4.1]{Fujii-Takahashi}.
\end{proof}
\end{proposition}


\begin{theorem}
\label{th-equilibrium-summary}
Let Assumptions~\ref{assumption-price-taker}, \ref{assumption-A} and \ref{assumption-B} be in force.
Then, there exists a unique market clearing equilibrium among the $N$ agents in which 
the equilibrium securities price processes are given by
\be
\vp_t=-\frac{1}{N}\sum_{i=1}^N Y_t^i, \quad t\in[0,T], 
\label{eq-equilibrium-price}
\ee
where $Y^i\in \mbb{S}^2(\mbb{F};\mbb{R}^n), ~1\leq i\leq N$ is the solution to the system of FBSDEs $(\ref{eq-fbsde-N})$. 
\begin{proof}
This is the direct consequence of  Theorems~\ref{th-existence-single}, \ref{th-N-equilibrium} and \ref{th-fbsde-N-existence}.
\end{proof}
\end{theorem}

\subsubsection*{Some economic observations}
Let us make several observations about the results in this section: 
First, it is well known that the backward component of the FBSDE arising from Pontryagin's maximum principle
represents the gradient of the value function. Since we are dealing with the minimization problem of the 
cost function, the result $(\ref{eq-equilibrium-price})$
implies that the prices of securities in the equilibrium are given by the (average of) {\it marginal utilities} of the $N$ agents.
This is quite reasonable from economics perspectives, since 
the amount of cash an agent can pay for some security
is naturally determined by how much gain he/she can expect in his/her utility function from 
possessing the security.

Second, since $\sigma_0^i$ and $\sigma_i$ are independent from the state variables,
Assumption~\ref{assumption-A} is enough to guarantee the existence of solution to $(\ref{eq-fbsde-N})$
for sufficiently small duration $T$.\footnote{See, for example, \cite[Theorem~1.1]{Delarue}}
Intuitively speaking, the convexity and monotonicity conditions in Assumption~\ref{assumption-B}
prevent the securities prices from blowing up (or crashing).
In particular, the monotonicity condition in Assumption~\ref{assumption-B} (ii)
implies that, in average,  the size of $(l_i)_i$ moves in the same direction as the securities' prices.
Recalling the explanation given in Section~\ref{subsec-model} on $l_i$, 
one can see that the demand of securities from the OTC clients 
tends to decrease as the prices of securities increase. 
Suppose otherwise the case. Then, when the prices go up, the OTC clients 
tend to request more securities from the registered financial firms, which reduces the storage level of securities among the agents.
Since each agent try to maintain the optimal level of his/her storage, the demand of the securities among the agents
also increases if the prices remain the same level. However, this would violate the market clearing condition.
In order to keep the demand unchanged (otherwise increase the supply), the prices of securities must go up higher.
It is natural to expect that this spiral effect pushes prices even higher to create a price bubble.
Although it is impossible to prove the one-to-one correspondence between the conditions in Assumption~\ref{assumption-B}
and the price bubble/crash, it looks at least reasonable that we need these conditions 
to guarantee the existence of equilibrium for an arbitrary time interval.
We may obtain more insights about this interesting  issue by investigating
an explicit solution possibly available in an appropriate linear-quadratic setup.

\section{Strong Convergence to the Mean-Field Limit}
\subsection{Convergence among the homogeneous agents}
Let us first consider the special case where the agents are homogeneous, i.e.
the coefficients and cost functions $(l_i, \sigma_i^0, \sigma_i, \ol{f}_i, \ol{g}_i)$ for each agent $1\leq i\leq N$
are equal to the common one $(l, \sigma^0, \sigma, \ol{f}, \ol{g})$.
In this case, the system of FBSDEs characterizing the  market clearing equilibrium among the $N$ price takers
becomes 
\bea
\begin{cases}
1\leq i \leq N,  \\
dX_t^i:=\Bigl\{ \wh{\alpha}\bigl(Y^i_t, -\mg{m}\bigl((Y_t^j)\bigr)\bigr)
+l\bigl(t, -\mg{m}\bigl((Y_t^j)\bigr), c_t^0,c_t^i\bigr)\Bigr\}dt+\sigma^0(t,c_t^0, c_t^i)dW_t^0+\sigma(t,c_t^0,c_t^i)dW_t^i, \\
dY_t^i=-\part_x \ol{f}\Bigl(t,X_t^i, -\mg{m}\bigl((Y_t^j)\bigr), c_t^0,c_t^i\Bigr)dt+
Z_t^{i,0}dW_t^0+\sum_{j=1}^N Z_t^{i,j}dW_t^j,   \\
\end{cases}
\label{eq-fbsde-N-Homo}
\eea
$t\in[0,T]$ with
\be
X_0^i=\xi^i,  \qquad Y_T^i=\frac{b}{1-b}\mg{m}\bigl((\part_x \ol{g}(X_T^j,c_T^0,c_T^j)\bigr)+\part_x \ol{g}(X_T^i,c_T^0,c_T^i).
\nn
\ee

Note that the $N$ sets of processes $\bigl(X^i, Y^i,.(Z^{i,j})_{j=0}^N\bigr), 1\leq i\leq N$, 
which consist of the unique solution of $(\ref{eq-fbsde-N-Homo})$,  are not independent  due to the 
interaction term $-\mg{m}((Y^i_t))$ arising from the price process. However, they are exchangeable 
since the interaction is symmetric and $(\xi^i)_{i=1}^N$ and $(c^i)_{i=1}^N$ are i.i.d. 
In particular, due to the exchangeability of  $(Y_t^i)_{i=1}^N$, 
De Finetti's theory of exchangeable sequence of random variables implies
\bea
\lim_{N\rightarrow \infty}\frac{1}{N}\sum_{i=1}^N Y_t^i=\mbb{E}\Bigl[Y_t^1|\bigcap_{k\geq 1}\sigma\{Y_t^j,j\geq k\}\Bigr]\quad{\rm a.s.}\nn
\eea
See for example \cite[Theorem 2.1]{Carmona-Delarue-2}. It also seems natural to expect that the tail $\sigma$-field is
reduced to $\ol{\calf}_t^0$ the $\sigma$-field generated by the common noise.

The above observation motivates us to investigate the following FBSDE of McKean-Vlasov typle for each agent $i\geq 1$.
\bea
\begin{cases}
dx_t^i=\Bigl(\wh{\alpha}\bigl(y_t^i, -\mbb{E}[y_t^i|\ol{\calf}_t^0]\bigr)+l\bigl(t, -\mbb{E}[y_t^i|\ol{\calf}_t^0],c_t^0,c_t^i\bigr)
\Bigr)dt+\sigma^0(t,c_t^0,c_t^i)dW_t^0+\sigma(t,c_t^0,c_t^i)dW_t^i,  \\
dy_t^i=-\part_x \ol{f}\bigl(t,x_t^i, -\mbb{E}[y_t^i|\ol{\calf}_t^0],c_t^0,c_t^i\bigr)dt+z_t^{i,0}dW_t^0+z_t^{i,i}dW_t^i, 
\end{cases}
\label{eq-fbsde-mfg}
\eea
for $t\in[0,T]$ with 
\be
x_0^i=\xi^i, \quad
y_T^i=\frac{b}{1-b}\mbb{E}\bigl[\part_x\ol{g}(x_T^i,c_T^0,c_T^i)|\ol{\calf}_T^0\bigr]+\part_x \ol{g}(x_T^i,c_T^0,c_T^i)~. \nn
\ee
This is the same equation studied in \cite{Fujii-Takahashi}.
In the following, we shall work on the bigger space $(\Omega^\infty, \calf^\infty, \mbb{P}^\infty)$
to support countably many $(\xi^i, W^i)_{i\geq 1}$ required to discuss the large population limit.
We introduce the following conditions.
Note that the conditions (ii) and (iii) are natural generalization of those of Assumption~\ref{assumption-B}
where they are given in terms of the empirical mean.

\begin{assumption}
\label{assumption-C}
(i)  $(b, \Lambda, l, \sigma^0, \sigma, \ol{f}, \ol{g})$ satisfies Assumptions~\ref{assumption-A}
and \ref{assumption-B} (i).  \\
(ii) For any $t\in[0,T]$, any random variables $x,x^\prime, c^0,c\in \mbb{L}^2(\calf^\infty;\mbb{R}^n)$
and any sub-$\sigma$ field $\calg\subset \calf^\infty$, the function $l$ satisfies the monotone condition
with some positive constants $\gamma^l>0$,
\be
\mbb{E}\Bigl[\bigl\langle l(t,\mbb{E}[x|\calg],c^0,c)-l(t,\mbb{E}[x\pp|\calg],c^0,c),x-x^\prime\bigr\rangle\Bigr]\geq \gamma^l 
\bold{1}_{\{L_\vp>0\}} \mbb{E}\Bigl[\mbb{E}\bigl[x-x^\prime|\calg\bigr]^2\Bigr]. \nn
\ee
(iii) There exists a strictly positive constant $\gamma$ satisfying $0<\gamma\leq \Bigl(\gamma^f-\frac{L_\vp^2}{4\gamma^l}\Bigr)\wedge \gamma^g$.
Moreover, for any random variables $x,x\pp,c^0,c\in \mbb{L}^2(\calf;\mbb{R}^n)$
and any sub-$\sigma$ field $\calg\subset \calf^\infty$, the function $\ol{g}$ satisfies
\bea
\frac{b}{1-b}\mbb{E}\Bigl[\bigl\langle \mbb{E}\bigl[\part_x\ol{g}(x,c^0,c)
-\part_x \ol{g}(x\pp,c^0,c)|\calg\bigr],x-x\pp\bigr\rangle\Bigr]\geq (\gamma-\gamma^g) \mbb{E}\bigl[|x-x^\prime|^2\bigr]. \nn
\eea
\end{assumption}

\begin{remark}[Lasry-Lions monotonicity]
The so-called Lasry-Lions monotonicity is a famous criterion for the uniqueness of the mean field games.
It dates back to their original papers~\cite{Lions-1, Lions-2, Lions-3}
and is defined as follows~\cite[Definition 3.28]{Carmona-Delarue-1}:
a real-valued function $h$ on $\mbb{R}^d\times \calp_2(\mbb{R}^d)$ is 
said to be monotone in the sense of Lasry and Lions, if, for all $\mu\in\calp_2(\mbb{R}^d)$,
the mapping $\mbb{R}^d\ni x\mapsto h(x,\mu)$ is at most quadratic 
growth,  and for all $\mu, \mu^\prime \in \calp_2(\mbb{R}^d)$, we have
\be
\int_{\mbb{R}^d}\bigl(h(x,\mu)-h(x,\mu^\prime)\bigr)d(\mu-\mu^\prime)(x)\geq 0. \nn
\ee
The uniqueness result in probabilistic settings is given by \cite[Theorem 3.29]{Carmona-Delarue-1}.
It says that there is at most one MFG equilibrium if the running as well as terminal {\it cost functions} satisfy 
Lasry-Lions monotonicity.
On the other hand, although the appearance is very similar (see, in particular, Assumption~\ref{assumption-C} (ii)), 
the relevant monotonicity used in the current paper has a different origin.
It  essentially corresponds to \cite[(H2.3)]{Peng-Wu}, which implies the monotonicity for the drift term in 
the product of the forward and backward components $\langle X_t, Y_t\rangle$ of the relevant  FBSDE.
The condition makes Banach's fixed  point theorem applicable for the  existence as well as the uniqueness of the solution,
and hence it is generally stronger than the former.
\end{remark}

We know the following result.
\begin{theorem}{(\cite[Theorem 4.2]{Fujii-Takahashi})}
\label{th-mfg-existence}
Let Assumption~\ref{assumption-C} be in force. Then,  for any $T>0$,  there exists
a unique strong solution $(x^i, y^i, z^{i,0}, z^{i,i})\in \mbb{S}^2(\mbb{F}^i; \mbb{R}^n)
\times \mbb{S}^2(\mbb{F}^i; \mbb{R}^n)\times \mbb{H}^2(\mbb{F}^i;\mbb{R}^{n\times d_0})\times 
\mbb{H}^2(\mbb{F}^i;\mbb{R}^{n\times d})$ to the FBSDE of conditional McKean-Vlasov type $(\ref{eq-fbsde-mfg})$
for each $i\geq 1$.
\end{theorem}

Note that  FBSDE $(\ref{eq-fbsde-mfg})$ is now decoupled 
for each $i\geq 1$.
In particular, for given $\ol{\calf}^0$ i.e. the common information, 
the solutions $(x^i, y^i, z^{i,0},z^{i,i}), i\geq 1$ are independently and identically distributed.
Because of this property, the quantities such as $\mbb{E}[y_t^i|\ol{\calf}_t^0]$ and $\mbb{E}[\part_x \ol{g}(x_T^i,c_T^0,c_T^i)|\ol{\calf}_T^0]$
are independent of the index $i$.

The FBSDE $(\ref{eq-fbsde-mfg})$ has been the major object of the analysis in the accompanying work \cite{Fujii-Takahashi},
in which we have found that 
the $\ol{\mbb{F}}^0$-progressively measurable process
\be
\vp_t^{\rm MFG}:=-\mbb{E}\bigl[y_t^i|\ol{\calf}^0_t] =-\mbb{E}\bigl[y_t^i|\ol{\calf}^0], \qquad t\in[0,T] \nn
\ee
provides a good approximate of the equilibrium market price if the agents have the common coefficients as in Assumption~\ref{assumption-C}.
In particular, we have proved in \cite[Theorem 5.1]{Fujii-Takahashi} that the process $\vp^{\rm MFG}$
achieves the asymptotic market clearing $(\ref{eq-asymptotic-clearing})$.
The goal of this section is to prove the 
strong convergence of the $N$-agent equilibrium given by Theorem~\ref{th-equilibrium-summary}
to the above mean-field limit when the agents are homogeneous.
Once this is done,  we can study the stability relation of the market price for the heterogeneous agents 
relative to the mean-field limit $\vp^{\rm MFG}$ with the help of Proposition~\ref{prop-N-stability}.

\subsubsection*{Conditional law on the product probability space}
Before going to the proof of convergence, let us briefly mention about
the conditional distribution on the product probability space.
For more details on the issue, see \cite[Section 2.1.3]{Carmona-Delarue-2}.
As mentioned before,  the analysis in this section is done in the filtered probability space
$(\Omega^\infty, \calf^\infty, \mbb{P}^\infty;\mbb{F}^\infty)$
which is the product of $(\ol{\Omega}^0, \ol{\calf}^0, \ol{\mbb{P}}^0;\ol{\mbb{F}}^0)$
and $(\ol{\Omega}^\infty, \ol{\calf}^\infty, \ol{\mbb{P}}^\infty;\ol{\mbb{F}}^\infty)$.
More precisely, $\Omega^\infty=\ol{\Omega}^0\times \ol{\Omega}^\infty$
and $(\calf^\infty, \mbb{P}^\infty)$ is the completion of $(\ol{\calf}^0\otimes \ol{\calf}^\infty, 
\ol{\mbb{P}}^0\otimes \ol{\mbb{P}}^\infty)$ and the filtration $\mbb{F}^\infty=(\calf_t^\infty)_{t\geq 0}$
is the complete and right continuous augmentation of $(\ol{\calf}_t^0\otimes \ol{\calf}_t^\infty)_{t\geq 0}$.
A generic element of $\Omega^\infty$ is denoted by $\omega=(\ol{\omega}^0, \ol{\omega}^\infty)$
where $\ol{\omega}^0\in\ol{\Omega}^0$ and $\ol{\omega}^\infty\in \ol{\Omega}^\infty$.
Due to the completion of $\ol{\calf}^0\otimes \ol{\calf}^\infty$,  the Fubini's theorem 
fails in general. However, it is known that the problem occurs in the exceptional (probability zero) event only. 
In particular, for any $\mbb{R}^n$-valued random variable $X$ on $(\Omega^\infty, \calf^\infty, \mbb{P}^\infty)$,
$X(\ol{\omega}^0,\cdot)$ is a random variable on $(\ol{\Omega}^\infty, \ol{\calf}^\infty, \ol{\mbb{P}}^\infty)$
for $\ol{\mbb{P}}^0$-a.s.  $\ol{\omega}^0\in \ol{\Omega}^0$.
By \cite[Lemma 2.4]{Carmona-Delarue-2},  the conditional law $\call(X|\ol{\calf}^0)$ of $X$ with given $\ol{\calf}^0$ satisfies
$\call(X|\ol{\calf}^0)(\ol{\omega}^0)=\call(X(\ol{\omega}^0, \cdot))$ for $\ol{\mbb{P}}^0$-a.s. $\ol{\omega}^0\in
\ol{\Omega}^0$.   Hence, one can actually 
define the conditional law by $\call(X(\ol{\omega}^0, \cdot))$ by assigning an arbitrary 
law for $\ol{\omega}^0$ in the null set in which $\call(X(\ol{\omega}^0, \cdot))$ is ill-defined.

Thanks to these properties under the simple product structure, one can extend 
the results of Glivenko-Cantelli theorem  \cite[Section 5.1.2.]{Carmona-Delarue-1} on the weak convergence of empirical measure 
to the situation in the presence of common noise. In fact, one can perform the same analysis with a fixed $\ol{\omega}^0$ (i.e. a fixed path 
of $W^0$) and then take the expectation with respect to $\ol{\mbb{P}}^0$ in the last step. 
This is the method actually used to derive the properties
of the conditional propagation of chaos \cite[Section 2.1.4, Theorem 2.12]{Carmona-Delarue-2}.

We now introduce the following measure arguments based on the solution to $(\ref{eq-fbsde-mfg})$;
\be
\begin{split}
&\ol{\mu}_t^N:=\frac{1}{N}\sum_{i=1}^N \del_{y_t^i}, 
\quad \call_t^0(y_t):=\call(y_t^1|\ol{\calf}_t^0), \quad t\in[0,T],  \\
& \ol{\mu}_g^N:=\frac{1}{N}\sum_{i=1}^N \del_{\part_x \ol{g}(x_T^i,c_T^0,c_T^i)}, \quad
\call_g^0:=\call\bigl(\part_x \ol{g}(x_T^1,c_T^0,c_T^1)|\ol{\calf}_T^0\bigr). 
\end{split}
\label{def-measures}
\ee
The next result is an important consequence of the above observations.

\begin{lemma}
\label{lemma-LN}
Let Assumption~\ref{assumption-C} be in force. 
Then we have
\be
\begin{split}
&\lim_{N\rightarrow\infty} \sup_{t\in[0,T]}\mbb{E}\Bigl[W_2(\ol{\mu}_t^N, \call_t^0(y_t))^2\Bigr]=0, \nn \\
&\lim_{N\rightarrow\infty} \mbb{E}\Bigl[W_2(\ol{\mu}_g^N, \call_g^0)^2\Bigr]=0.\nn
\end{split}
\ee
Moreover, if there exist some positive constants $\Gamma$ and $\Gamma_g$ such that
$\sup_{t\in[0,T]}\mbb{E}[|y_t^i|^q]^\frac{1}{q}\leq \Gamma$ and $\mbb{E}\bigl[|\part_x \ol{g}(x_T^i,c_T^0,c_T^i)|^q\bigr]^\frac{1}{q}
\leq \Gamma_g$ for some $q>4$, 
then there exists some constant $C$ independent of $N$ such that
\bea
&&\sup_{t\in[0,T]}\mbb{E}\Bigl[W_2(\ol{\mu}_t^N, \call_t^0(y_t))^2\Bigr]\leq C\Gamma^2\ep_N,  \nn \\
&&\mbb{E}\Bigl[W_2(\ol{\mu}_g^N, \call_g^0)^2\Bigr]\leq C\Gamma_g^2 \ep_N, \nn
\eea
where $\ep_N:=N^{-2/\max(n,4)}\bigl(1+\log(N)\bold{1}_{N=4}\bigr)$.
\begin{proof}
In the first assertion, the claim for $W_2(\ol{\mu}_t^N, \call_t^0(y_t))$ is proved in \cite[Theorem~5.1]{Fujii-Takahashi},
which is the straightforward generalization of \cite[Theorem 2.12]{Carmona-Delarue-2}.
For completeness, we give the details below.

Since $(y_t^i)_{i\geq 1}$ are $\ol{\calf}^0_t$-conditionally independently and identically distributed,
the Glivenko-Cantelli theorem implies 
\be
\mbb{P}\Bigl(\Bigl\{\lim_{N\rightarrow \infty}\mbb{E}\Bigl[W_2(\ol{\mu}_t^N, \call_t^0(y_t))^2|\ol{\calf}^0_t\Bigr]=0\Bigr\}\Bigr)=1. \nn
\ee
Since we have
\be
\begin{split}
&\mbb{E}\Bigl[W_2(\ol{\mu}_t^N, \call_t^0(y_t))^2|\ol{\calf}^0_t\Bigr]=\mbb{E}\Bigl[W_2\Bigl(\frac{1}{N}\sum_{i=1}^N \del_{y_t^i}, \call(y_t^1|\ol{\calf}_t^0)\Bigr)^2\Bigr|\ol{\calf}_t^0\Bigr] \\
&\quad \leq\frac{2}{N}\sum_{i=1}^N \mbb{E}\bigl[|y_t^i|^2|\ol{\calf}_t^0\bigr]+2\mbb{E}\bigl[|y_t^1|^2|\ol{\calf}_t^0\bigr]
=4\mbb{E}\bigl[|y_t^1|^2|\ol{\calf}_t^0\bigr], 
\end{split}\nn
\ee
and know that $y^1\in \mbb{S}^2$,  we can apply the dominated convergence theorem to conclude that
the pointwise convergence holds:
\bea
\lim_{N\rightarrow \infty}\mbb{E}\Bigl[W_2\Bigl(\ol{\mu}_t^N, \call^0(y_t)\Bigr)^2\Bigr]=0~.
\label{pointwise-conv}
\eea
We are now going to show that the set of functions, $(f_N)_{N\in \mbb{N}}$ defined by
\bea
[0,T]\ni t\mapsto f_N(t):=\mbb{E}\bigl[W_2\bigl(\ol{\mu}_t^N, \call^0(y_t)\bigr)^2\bigr]\in \mbb{R} \nn
\eea
are precompact in the space $\calc([0,T]; \mbb{R})$ endowed with the topology of uniform convergence.
In fact, uniformly in $N$,
\bea
\sup_{t\in[0,T]}|f_N(t)|\leq 4\sup_{t\in[0,T]}\mbb{E}\bigl[|y_t^1|^2\bigr]<\infty.
\label{fN-inequality}
\eea
Moreover,  for any $0\leq t,s\leq T$,  Cauchy-Schwarz, $(\ref{fN-inequality})$ and the triangular inequalities give,
with some constant $C$ independent of $N$, 
\be
\begin{split}
&|f_N(t)-f_N(s)|\\
&\quad \leq \mbb{E}\Bigl[\Bigl(W_2(\ol{\mu}_t^N,\call^0(y_t))+W_2(\ol{\mu}^N_s,\call^0(y_s)\Bigr)^2\Bigr]^\frac{1}{2}\mbb{E}\Bigl[\Bigl(W_2(\ol{\mu}^N_t,\call^0(y_t))-W_2(\ol{\mu}^N_s,\call^0(y_s))\Bigr)^2\Bigr]^\frac{1}{2} \\
&\quad \leq C\mbb{E}\Bigl[\Bigl(W_2(\ol{\mu}_t^N,\call^0(y_t))-W_2(\ol{\mu}_s^N,\call^0(y_s))\Bigr)^2\Bigr]^\frac{1}{2} \leq 
C\mbb{E}\Bigl[ W_2(\ol{\mu}^N_t,\ol{\mu}^N_s)^2+W_2(\call^0(y_t),\call^0(y_s))^2\Bigr]^\frac{1}{2}~ \\
&\quad \leq C\mbb{E}\Bigl[ \frac{1}{N}\sum_{i=1}^N |y_t^i-y_s^i|^2+|y_t^1-y_s^1|^2\Bigr]^\frac{1}{2} 
\leq C\mbb{E}\bigl[|y_t^1-y_s^1|^2\bigr]^\frac{1}{2},
\end{split} \nn
\ee
where we have used the fact that $(y^i)_{i\geq 1}$ are conditionally i.i.d at the last inequality.

Since $(y_t^1)_{t\in[0,T]}$ is a continuous process, the above estimate tells that
$(f_N)_{N\in \mbb{N}}$ is equicontinuous, which is also uniformly equicontinuous since we are working on the finite interval.
Now, Arzela-Ascoli theorem implies the desired precompactness.  
Combining with the pointwise convergence $(\ref{pointwise-conv})$,  we thus conclude
$\lim_{N\rightarrow \infty}\sup_{t\in[0,T]}\mbb{E}\bigl[W_2\bigl(\ol{\mu}_t^N, \call^0(y_t)\bigr)^2\bigr]=0$.
Since $\Bigl(\part_x\ol{g}(x_T^i,c_T^0,c_T^i)\Bigr)_{i=1}^N$ are $\ol{\calf}_T^0$-conditionally 
i.i.d. square integrable random variables, the claim for the $W_2(\ol{\mu}_g^N, \call_g^0)$ is established
in the same way. Since the time $T$ is fixed, the continuity property used for $W_2(\ol{\mu}_t^N, \call_t^0(y_t))$
is unnecessary. 

The second assertion of the non-asymptotic estimate on the convergence order in $N$ is the direct consequence of 
\cite[Theorem 5.8, Remark 5.9]{Carmona-Delarue-1} as well as the previous observations on the conditional law.
\end{proof}
\end{lemma}

\begin{remark}
The integrability condition for the second assertion of the last lemma is satisfied if,   for some $q>4$,  
$\xi^i\in \mbb{L}^q(\ol{\calf}_0^i;\mbb{R}^n)$ for every $i\geq 1$,  
and $c_T^j\in \mbb{L}^q(\ol{\calf}_T^j;\mbb{R}^n)$, $\mbb{E}\Bigl[\Bigl(\int_0^T |c_t^j|^2dt\Bigr)^{q/2}\Bigr]<\infty$
for every $j\geq 0$. See, for example,  the discussions in
\cite{Li-Wei} or \cite[Theorem 4.1]{J-Yong}
\end{remark}

The next theorem is the second main result of the paper.
\begin{theorem}
\label{th-strong-convergence}
Suppose that 
Assumption~\ref{assumption-C} and also the conditions $(ii)$ and $(iii)$ of Assumption~\ref{assumption-B} are satisfied.
Let $(X^i, Y^i,Z^{i,0},(Z^{i,j})_{j=1}^N)_{i=1}^N$  and $(x^i, y^i, z^{i,0}, z^{i,i}), ~1\leq i\leq N$
denote the unique strong solution to the $N$-coupled system of FBSDEs $(\ref{eq-fbsde-N-Homo})$
 and that of the FBSDE of conditional McKean-Vlasov type
$(\ref{eq-fbsde-mfg})$ with $1\leq i\leq N$, respectively.
Then, there exists some $N$-independent constant $C$ such that
\bea
&&\mbb{E}\Bigl[\sup_{t\in[0,T]}|\Del X_t^i|^2+\sup_{t\in[0,T]}|\Del Y_t^i|^2+\int_0^T
\bigl(|\Del Z_t^{i,0}|^2+\sum_{j=1}^N |\Del Z_t^{i,j}|^2\bigr)dt\Bigr]\nn \\
&&\quad \leq C\mbb{E}\Bigl[W_2(\ol{\mu}_g^N, \call_g^0)^2+\int_0^T W_2(\ol{\mu}_t^N, \call_t^0(y_t))^2dt\Bigr], 
\label{eq-conv-result}
\eea
where $\Del X^i:=X^i-x^i$, $\Del Y^i:=Y^i-y^i$, $\Del Z^{i,0}:=Z^{i,0}-z^{i,0}$ and
$\Del Z^{i,j}:=Z^{i,j}-\del_{i,j}z^{i,i}$.

\begin{proof}
Using the notations in $(\ref{def-BFG})$, we have for each $1\leq i\leq N$, 
\bea
\begin{cases}
d\Del X_t^i=\Bigl(B(t,Y_t^i,\mu_t^N)-B(t,y_t^i,\call_t^0(y_t))\Bigr)dt, \nn \\
d\Del Y_t^i=\Bigl(F(t,X_t^i,\mu_t^N)-F(t,x_t^i,\call_t^0(y_t))\Bigr)dt+\Del Z_t^{i,0}dW_t^0+
\sum_{j=1}^N \Del Z_t^{i,j}dW_t^j, 
\end{cases} \nn
\eea
for $t\in[0,T]$ where $\mu_t^N:=\frac{1}{N}\sum_{i=1}^N \del_{Y_t^i}$ is the empirical measure.
To lighten the expression, we omit the arguments $(c^0_t,c^i_t)$, which does not play an important role for the 
stability analysis below.
\\
{\it First Step}: It is important to notice the inequality
\bea
\Bigl|\frac{1}{N}\sum_{i=1}^N y_t^i-\mbb{E}[y_t^i|\ol{\calf}_t^0]\Bigr|\leq W_2\bigl(\ol{\mu}_t^N, \call^0_t(y_t)\bigr).\nn
\eea
This is understood as follows; for an arbitrary pair $\mu,\nu\in \calp_2(\mbb{R}^n)$, we have
\bea
\Bigl|\int_{\mbb{R}^n} x \mu(dx)-\int_{\mbb{R}^n} y \nu(dy)\Bigr|=\Bigl| \int_{\mbb{R}^n\times \mbb{R}^n}(x-y)\pi(dx,dy)\Bigr|\leq 
\int_{\mbb{R}^n\times \mbb{R}^n}|x-y|\pi(dx,dy), 
\label{wasser-ineq}
\eea
for any coupling $\pi\in \Pi_2(\mu,\nu)$ with marginals $\mu$ and $\nu$. Taking the infimum over $\pi\in \Pi_2(\mu,\nu)$,
we get 
\bea
|m(\mu)-m(\nu)|\leq W_1(\mu,\nu)\leq W_2(\mu, \nu),  \nn
\eea
by the definition of the Wasserstein distance $(\ref{def-W})$.
With $\mu=\ol{\mu}_t^N$ and $\nu=\call^0_t(y_t)$, we obtain the desired inequality.
From Assumption~\ref{assumption-B} (i) 
and the above observation,  one can see that $B$ and $F$ are both Lipschitz continuous in their 
measure argument with respect to the $W_2$-distance.

The above observation combined with $(\ref{ineq-B})$, we get
\be
\begin{split}
&\sum_{i=1}^N \langle B(t,Y_t^i, \mu_t^N)-B(t,y_t^i, \call_t^0(y_t)), \Del Y_t^i\rangle  \\
&\quad =\sum_{i=1}^N \langle B(t,Y_t^i, \mu_t^N)-B(t,y_t^i, \ol{\mu}_t^N), \Del Y_t^i\rangle+
\sum_{i=1}^N \langle B(t,y_t^i, \ol{\mu}_t^N)-B(t,y_t^i, \call_t^0(y_t)), \Del Y_t^i\rangle  \\
&\quad \leq -N\gamma^l \bold{1}_{\{L_\vp>0\}}\bigl|\mg{m}((\Del Y_t^i)) \bigr|^2+
C\sum_{i=1}^N W_2(\ol{\mu}_t^N, \call_t^0(y_t))|\Del Y_t^i|. 
\end{split}\nn
\ee
Using now $(\ref{ineq-F})$, similar procedures yield
\be
\begin{split}
&\sum_{i=1}^N \langle F(t,X_t^i, \mu_t^N)-F(t,x_t^i,\call_t^0(y_t)), \Del X_t^i\rangle  \\
&\quad \leq -\Bigl(\gamma^f-\frac{L_\vp^2}{4\gamma^l}\Bigr)\sum_{i=1}^N |\Del X_t^i|^2+
N\gamma^l \bold{1}_{\{L_\vp>0\}}\bigl|\mg{m}((\Del Y_t^i))\bigr|^2+C\sum_{i=1}^N W_2(\ol{\mu}_t^N, \call_t^0(y_t))|\Del X_t^i|.
\end{split}\nn
\ee
Since $\Del X_0^i=0$ for every $i$, by simple application of \Ito-formula and the above estimates, 
we obtain
\be
\begin{split}
\sum_{i=1}^N \mbb{E}\Bigl[\bigl\langle \Del X_T^i, \Del Y_T^i\bigr\rangle\Bigr]
&\leq -\Bigl(\gamma^f-\frac{L_\vp^2}{4\gamma^l}\Bigr)\sum_{i=1}^N \mbb{E}\int_0^T |\Del X_t^i|^2 dt  \\
&+C\sum_{i=1}^N \mbb{E}\int_0^T W_2(\ol{\mu}_t^N, \call_t^0(y_t))\bigl(|\Del X_t^i|+|\Del Y_t^i|\bigr)dt~.
\end{split}
\label{eq-conv-above}
\ee
On the other hand, with $\mu_g^N:=\frac{1}{N}\sum_{i=1}^N \part_x \ol{g}(X_T^i, c_T^0,c_T^i)$, 
we have from the terminal condition
\be
\begin{split}
&\sum_{i=1}^N \langle G(\mu_g^N, X_T^i)-G(\call_g^0,x_T^i),\Del X_T^i\rangle  \\
&=\sum_{i=1}^N \langle G(\mu_g^N, X_T^i)-G(\ol{\mu}_g^N, x_T^i),\Del X_T^i\rangle
+\sum_{i=1}^N \langle G(\ol{\mu}_g^N, x_T^i)-G(\call_g^0, x_T^i),\Del X_T^i\rangle, 
\end{split}
\nn
\ee
where we have omitted $(c^0_t, c_T^i)$ to lighten the notation.
Now applying the inequality $(\ref{ineq-G})$ to the first term, we get
\be
\sum_{i=1}^N \mbb{E}\Bigl[\bigl\langle \Del Y_T^i, \Del X_T^i\bigr\rangle\Bigr] \geq \gamma \sum_{i=1}^N \mbb{E}\bigl[|\Del X_T^i|^2\bigr]-
\frac{b}{1-b}\sum_{i=1}^N \mbb{E}
\Bigl[W_2(\ol{\mu}_g^N, \call_g^0)|\Del X_T^i|\Bigr].
\label{eq-conv-below}
\ee
Combining the two estimates $(\ref{eq-conv-above})$ and $(\ref{eq-conv-below})$, we have
\be
\begin{split}
\sum_{i=1}^N \mbb{E}\Bigl[|\Del X_T^i|^2+\int_0^T |\Del X_t^i|^2 dt\Bigr]
&\leq C\sum_{i=1}^N \mbb{E}\int_0^T W_2(\ol{\mu}_t^N,\call^0_t(y_t))\big[|\Del X_t^i|+|\Del Y_t^i|\bigr]dt \\
&+C\sum_{i=1}^N\mbb{E}\Bigl[W_2(\ol{\mu}_g^N, \call_g^0)|\Del X_T^i|\Bigr], 
\end{split} \nn
\ee
where the constant $C$ now depends on $\gamma, b$ but not on $N$. 
Since the random variables such as $\Del X^i, \Del Y^i$ 
have the same distributions for all $i$ due to the common coefficient functions, 
the assumptions on $\xi^i$ and $c^i$  and the structure of the probability space, we have
\be
\begin{split}
\mbb{E}\Bigl[|\Del X_T^i|^2+\int_0^T |\Del X_t^i|^2 dt\Bigr]
&\leq C \mbb{E}\int_0^T W_2(\ol{\mu}_t^N,\call^0_t(y_t))\big[|\Del X_t^i|+|\Del Y_t^i|\bigr]dt \\
&+C\mbb{E}\Bigl[W_2(\ol{\mu}_g^N, \call_g^0)|\Del X_T^i|\Bigr], 
\end{split} \nn
\ee
for every $1\leq i\leq N$. Now, from Young's inequality,  we obtain
\be
\begin{split}
\mbb{E}\Bigl[|\Del X_T^i|^2+\int_0^T |\Del X_t^i|^2 dt\Bigr]&\leq C\mbb{E}\int_0^T 
\bigl[W_2(\ol{\mu}_t^N, \call_t^0(y_t))^2+W_2(\ol{\mu}_t^N, \call_t^0(y_t))|\Del Y_t^i|\bigr]dt \\
&+C\mbb{E}\Bigl[W_2(\ol{\mu}_g^N, \call_g^0)^2\Bigr], 
\end{split}
\label{eq-1st-final}
\ee
for every $1\leq i\leq N$.
\\
{\it Second Step}:
A simple application of \Ito-formula to $|\Del Y_t^i|^2$ gives, for any $t\in[0,T]$, 
\be
\begin{split}
&\mbb{E}\Bigl[|\Del Y_t^i|^2+\int_t^T \bigl(|\Del Z_s^{i,0}|^2+\sum_{j=1}^N |\Del Z_s^{i,j}|^2\bigr)ds\Bigr] \\
&\qquad=\mbb{E}\Bigl[|\Del Y_T^i|^2-2\int_t^T \langle F(s,X_s^i, \mu_s^N)-F(s,x_s^i,\call_s^0(y_s)),\Del Y_s^i\rangle ds\Bigr]~.
\end{split}
\label{eq-square-Ito}
\ee
Note that, from Assumption~\ref{assumption-A} (iii) and the estimate $(\ref{wasser-ineq})$, 
\be
\begin{split}
|\Del Y_T^i|&\leq |G(\mu_g^N, X_T^i)-G(\ol{\mu}_g^N, x_T^i)|+|G(\ol{\mu}_g^N, x_T^i)-G(\call_g^0,x_T^i)| \\
&\leq  C\Bigl( \mg{m}((|\Del X_T^j|))+|\Del X_T^i|+W_2(\ol{\mu}_g^N, \call_g^0)\Bigr). 
\end{split} \nn
\ee
Using this estimate and the exchangeability of variables, we obtain from $(\ref{eq-square-Ito})$ that
\be
\begin{split}
&\mbb{E}\Bigl[|\Del Y_t^i|^2+\int_t^T \bigl(|\Del Z_s^{i,0}|^2+\sum_{j=1}^N |\Del Z_s^{i,j}|^2\bigr)ds\Bigr]  \\
&\quad \leq C\mbb{E}\Bigl[|\Del X_T^i|^2+W_2(\ol{\mu}_g^N, \call_g^0)^2\Bigr]
+C\mbb{E}\int_t^T \bigl[ |\Del X_s^i|+W_2(\mu_s^N, \call_s^0(y_s))\bigr]|\Del Y_s^i|ds  \\
&\quad \leq C\mbb{E}\Bigl[|\Del X_T^i|^2+W_2(\ol{\mu}_g^N, \call_g^0)^2\Bigr]
+C\mbb{E}\int_0^T \bigl(|\Del X_s^i|^2+W_2(\ol{\mu}_s^N, \call_s^0(y_s))^2\bigr)ds+
C\mbb{E}\int_t^T |\Del Y_s^i|^2 ds,
\end{split} \nn
\ee
for every $1\leq i\leq N$. Here, we have used the triangle inequality w.r.t. the Wasserstein distance $W_2$ and the fact that
\bea
\mbb{E}\Bigl[W_2(\mu_s^N, \ol{\mu}_s^N)^2\Bigr]\leq \mbb{E}\Bigl[\frac{1}{N}\sum_{j=1}^N |Y_s^j-y_s^j|^2\Bigr]=
\mbb{E}|\Del Y_s^i|^2.
\label{eq-dummy}
\eea
By applying the backward Gronwall's inequality and the estimate $(\ref{eq-1st-final})$, we get
\be
\begin{split}
&\sup_{t\in[0,T]}\mbb{E}\Bigl[|\Del Y_t^i|^2\Bigr]+\mbb{E}\int_0^T \bigl(|\Del Z_t^{i,0}|^2+\sum_{j=1}^N |\Del Z_t^{i,j}|^2\bigr)dt\nn \\
&\quad \leq C\mbb{E}\Bigl[W_2(\ol{\mu}_g^N,\call_g^0)^2+\int_0^T\bigl( W_2(\ol{\mu}_t^N, \call_t^0(y_t))^2+W_2(\ol{\mu}_t^N, \call_t^0(y_t))
|\Del Y_t^i|\bigr)dt\Bigr].
\end{split}\nn
\ee
Using Young's inequality, we obtain
\be
\begin{split}
&\sup_{t\in[0,T]}\mbb{E}\Bigl[|\Del Y_t^i|^2\Bigr]+\mbb{E}\int_0^T \bigl(|\Del Z_t^{i,0}|^2+\sum_{j=1}^N |\Del Z_t^{i,j}|^2\bigr)dt \\
&\hspace{25mm} \leq C\mbb{E}\Bigl[W_2(\ol{\mu}_g^N,\call_g^0)^2+\int_0^TW_2(\ol{\mu}_t^N, \call_t^0(y_t))^2dt\Bigr], 
\end{split}
\label{eq-2nd-final-1}
\ee
from which and $(\ref{eq-1st-final})$, we also have
\be
\mbb{E}\Bigl[|\Del X_T^i|^2+\int_0^T |\Del X_t^i|^2 dt\Bigr]\leq C\mbb{E}\Bigl[W_2(\ol{\mu}_g^N,\call_g^0)^2+\int_0^TW_2(\ol{\mu}_t^N, 
\call_t^0(y_t))^2dt\Bigr]. 
\label{eq-2nd-final-2}
\ee
The inequality $(\ref{eq-conv-result})$ now easily follows from $(\ref{eq-2nd-final-1})$, $(\ref{eq-2nd-final-2})$ 
and the standard application of the Burkholder-Davis-Gundy inequality.
\end{proof}
\end{theorem}

Combined with Lemma~\ref{lemma-LN}, Theorem~\ref{th-strong-convergence}
implies that the $i$th component $(X^i, Y^i, Z^{i,0}, Z^{i,i})$ of the solution of  $(\ref{eq-fbsde-N-Homo})$
converges strongly to the solution $(x^i, y^i, z^{i,0}, z^{i,i})$ of  
$(\ref{eq-fbsde-mfg})$, where  $Z^{i,j}, ~j\neq i$ converge to zero.
Note that  $(\ref{eq-fbsde-mfg})$ is equivalent to $(\ref{eq-fbsde-single})$
in  the setup with $\vp_t=-\mbb{E}[y^1_t|\ol{\calf}^0_t], ~t\geq 0$, which is adapted to $\ol{\mbb{F}}^0$
the filtration generated by the common noise. 
Therefore, in the large population limit,  the optimal strategy for each agent $i$ is unchanged even if we restrict the space
of his/her admissible strategies to $\mbb{A}^i:=\mbb{H}^2(\mbb{F}^i;\mbb{R}^n)$.
Recalling that  $\mbb{F}^i$ is the product of $\ol{\mbb{F}}^0$ and $\ol{\mbb{F}}^i$,
the idiosyncratic information for the other agents $(\ol{\mbb{F}}^j,j\neq i)$
is not required anymore. As a result,  there is no need to 
impose the perfect information assumption  as announced in Remark~\ref{remark-info}.

\begin{remark}
\label{remark-tangpi}
Although it is for a specific economic model, 
let us emphasize that the proof of convergence based on the monotonicity conditions for an arbitrary time interval
was given in the first time in our preprint~\cite{FT-finite-agent} (Oct. 2020), which is the first version in arXiv
of the current manuscript. 
Although one can find related results on backward propagation of chaos in the recent work~\cite{Lauriere} by Lauri\`{e}re and Tangpi,
the proof given in their first version (Apr. 2020)  in arXiv adopted 
a quite different approach, where the short-term estimates were  sticked together (Theorem~12).
In the latest version of their manuscript,  which is the second version in arXiv, 
the corresponding result in Theorem~14 is now restricted to the case of sufficiently small $T$.
The new result in Theorem~18,  that proves the convergence for general $T$, 
is now based on the similar monotonicity conditions as ours.
However this version of the paper was published at (Apr. 2021), i.e. after the publication  of our manuscript in arXiv. 
Therefore, as the timeline  suggests,
the direct application of monotonicity conditions related to those in \cite{Peng-Wu} is our original and given independently from their results.
As for the difference from their latest version, our monotonicity directly involves the measure argument and also the common noise,
which is not the case in their work.
\end{remark}

Under the conditions used in Theorem~\ref{th-strong-convergence}, the market clearing price 
for the homogeneous agents is given by 
\bea
\vp_t^{\rm Ho}:=-\frac{1}{N}\sum_{i=1}^N Y_t^i, \quad t\in [0,T], \nn
\eea
where $(Y^i)_{i=1}^N$ is the solution to the $N$-coupled system of FBSDEs $(\ref{eq-fbsde-N-Homo})$.
On the other hand, the price process in the mean-field limit is 
given by
\bea
\vp_t^{\rm MFG}:=-\mbb{E}\bigl[y_t^1|\ol{\calf}_t^0\bigr], \quad t\in[0,T],
\eea
which is proven to clear the market asymptotically in the large population limit~\cite{Fujii-Takahashi}.

\begin{corollary}
Let Assumption~\ref{assumption-C} and also the conditions $(ii)$ and $(iii)$ of Assumption~\ref{assumption-B}
 be in force. With the above notations, we have 
\bea
&&\sup_{t\in[0,T]}\mbb{E}\Bigl[|\vp_t^{\rm Ho}-\vp_t^{\rm MFG}|^2\Bigr]
+\mbb{E}\Bigl[\sup_{t\in[0,T]}\bigl| \mbb{E}[\vp_t^{\rm Ho}|\ol{\calf}_t^0]-\vp_t^{\rm MFG}\bigr|^2\Bigr]\nn \\
&&\qquad \leq C\Bigl(\sup_{t\in[0,T]}\mbb{E}\Bigl[W_2(\ol{\mu}_t^N,\call^0_t(y_t))^2\Bigr]+\mbb{E}\Bigl[W_2(\ol{\mu}_g^N, \call_g^0)^2\Bigr]\Bigr)
~,\nn
\eea
where $C$ is some $N$-independent constant.
\begin{proof}
Using $(\ref{wasser-ineq})$, we have
\be
\begin{split}
\bigl|\vp_t^{\rm Ho}-\vp_t^{\rm MFG}\bigr|^2&=\bigl|m(\mu_t^N)-m(\call^0_t(y_t))\bigr|^2\nn \\
&\leq W_2(\mu_t^N, \call_t^0(y_t))^2\leq 2W_2(\mu_t^N,\ol{\mu}_t^N)^2+2W_2(\ol{\mu}_t^N, \call_t^0(y_t))^2.
\end{split}
\nn
\ee
The desired estimate for the first term now follows from $(\ref{eq-dummy})$.

Note that for any constants $a_i\in \mbb{R}, 1\leq i\leq N$, we have $(\sum_{i=1}^N a_i)^2\leq N\sum_{i=1}^N |a_i|^2$.
Hence
\be
\begin{split}
&\mbb{E}\Bigl[\sup_{t\in[0,T]}\bigl| \mbb{E}[\vp_t^{\rm Ho}|\ol{\calf}_t^0]-\vp_t^{\rm MFG}\bigr|^2\Bigr]
=\mbb{E}\Bigl[\sup_{t\in[0,T]}\Bigl|\frac{1}{N}\sum_{i=1}^N \mbb{E}\bigl[Y^i_t-y_t^i|\ol{\calf}_t^0\bigr]\Bigr|^2\Bigr]\\
&\quad \leq \mbb{E}\Bigl[\sup_{t\in[0,T]}\frac{1}{N}\sum_{i=1}^N \mbb{E}\bigl[|Y^i_t-y_t^i|^2 |\ol{\calf}_t^0\bigr]\Bigr]\\
&\quad \leq \mbb{E}\Bigl[\mbb{E}\Bigl[ \sup_{t\in[0,T]}\sum_{i=1}^N \frac{1}{N} |Y_t^i-y_t^i|^2 |\ol{\calf}^0\Bigr]\Bigr]
=\mbb{E}\bigl[\sup_{t\in[0,T]}|\Del Y_t^1|^2\bigr]. 
\end{split} \nn
\ee
In the second equality, we used the fact  that $\ol{\mbb{F}}^0$ is generated by the Brownian motion $W^0$ 
and hence the additional information contained in $\ol{\calf}_s^0, s\geq t$ does not  affect the expectation value
of $\calf^\infty_t$-measurable random variables.
Thus, we have 
\be
\mbb{E}\bigl[|Y_t^i-y_t^i|^2|\ol{\calf}_t^0\bigr]=\mbb{E}\bigl[|Y_t^i-y_t^i|^2|\ol{\calf}^0\bigr] \nn
\ee
for any $t\in[0,T]$.  Using the exchangeability of $(Y^i, y^i)$ and the result of Theorem~\ref{th-strong-convergence},
we obtains the desired estimate for the second term.
\end{proof}
\end{corollary}

\subsection*{Stability of the market price for the heterogeneous agents}
Suppose that the $N$ agents have the common discount parameter $b$
and the common rate of the trading fee $\Lambda$ to be paid to the securities exchange.
Instead of the homogeneous agents, we now consider the case where 
the agents have different cost functions and different order-flow from their clients;
$(l_i,\sigma_i^0, \sigma_i, \ol{f}_i, \ol{g}_i), 1\leq i\leq N$.
Except the overall structure assumed in $(\ref{eq-cost-functions})$,
the cost functions $(\ol{f}_i, \ol{g}_i)$ can be changed freely as long as they satisfy the convexity 
as well as the monotonicity conditions uniformly.
This is clear contrast to the existing literature where one can change only the risk-tolerance coefficient,
such as  $\gamma^i$ in $\displaystyle \exp\Bigl(-\frac{x}{\gamma^i}\Bigr)$ of the exponential utility function, for example.

\begin{proposition}
\label{prop-hetero-conv}
Assume that the coefficients $(b, \Lambda, l_i,\sigma_i^0,\sigma_i,\ol{f}_i,\ol{g}_i)_{i=1}^N$
satisfy  Assumptions~\ref{assumption-A} and \ref{assumption-B}, 
and that $(l,\sigma^0,\sigma, \ol{f},\ol{g})$ satisfy Assumption~\ref{assumption-C}
and the conditions (ii), (iii) of Assumption~\ref{assumption-B}.
Let us denote by
$\bigl(\ch{X}^i, \ch{Y}^i, \ch{Z}^{i,0}, (\ch{Z}^{i,j})_{j=1}^N\bigr)_{i=1}^N$, $\bigl(X^i, Y^i, Z^{i,0}, (Z^{i,j})_{j=1}^N\bigr)$
and $(x^i, y^i, z^{i,0}, z^{i,i}), i\geq 1$ the unique solution to
$(\ref{eq-fbsde-N})$, $(\ref{eq-fbsde-N-Homo})$ and $(\ref{eq-fbsde-mfg})$, respectively.
Then there exists some $N$ independent constant $C$ such that
\be
\begin{split}
&\sum_{i=1}^N \mbb{E}\Bigl[\sup_{t\in[0,T]}|\Del \ch{X}_t^i|^2+\sup_{t\in[0,T]}|\Del \ch{Y}_t^i|^2+
\int_0^T \bigl(|\Del \ch{Z}_t^{i,0}|^2+\sum_{j=1}^N |\Del \ch{Z}_t^{i,j}|^2\bigr)dt\Bigr] \\
&\quad \leq  C N\mbb{E}\Bigl[W_2(\ol{\mu}_g^N, \call_g^0)^2+\int_0^T W_2(\ol{\mu}_t^N, \call_t^0(y_t))^2dt\Bigr] \\
&\quad + C\sum_{i=1}^N \mbb{E}\Bigl[|\del G_i|^2+\int_0^T 
\bigl(|\del F_i(t)|^2+|\del B_i(t)|^2+|\del \sigma_i^0(t)|^2+|\del \sigma_i(t)|^2\bigr)dt\Bigr]~, 
\end{split} \nn
\ee
where $\Del \ch{X}^i:=\ch{X}^i-x^i$, $\Del \ch{Y}^i:=\ch{Y}^i-y^i$, 
$\Del \ch{Z}^{i,0}:=\ch{Z}^{i,0}-z^{i,0}$, $\Del \ch{Z}^{i,j}=\ch{Z}^{i,j}-\del_{i,j}z^{i,i}$ and 
\be
\begin{split}
&\del B_i(t):=\bigl(l_i-l)\Bigl(t, Y^i_t, \vp_t^{\rm Ho}, c_t^0,c_t^i\Bigr), 
\qquad \del F_i(t):=-\bigl(\part_x \ol{f}_i-\part_x \ol{f}\bigr)\Bigl(t,X_t^i, \vp_t^{\rm Ho}, c_t^0,c_t^i\Bigr),  \\
&\del G_i:=\frac{b}{1-b}\sum_{j=1}^N \bigl(\part_x \ol{g}_j-\part_x \ol{g}\bigr)(X_T^j, c_T^0,c_T^j)
+(\part_x \ol{g}_i-\part_x \ol{g})(X_T^i, c_T^0,c_T^i).  \\
&(\del \sigma_i^0,\del \sigma_i)(t):=\bigl((\sigma_i^0, \sigma_i)(t,c_t^0,c_t^i)-
(\sigma^{0}, \sigma)(t,c_t^{0}, c_t^{i})\bigr). 
\end{split} \nn
\ee
\begin{proof}
This is the direct consequence of Proposition~\ref{prop-N-stability}
and Theorem~\ref{th-strong-convergence}.
\end{proof}
\end{proposition}

From Theorem~\ref{th-equilibrium-summary}
we know that the market clearing price among the $N$ heterogeneous agents 
is given by
\bea
\vp_t^{\rm He}:=-\frac{1}{N}\sum_{i=1}^N\ch{Y}_t^i, \quad t\in[0,T].\nn
\eea
The next corollary gives the stability result of the market price around the mean-field limit.

\begin{corollary}
\label{co-price-stability}
Under the assumptions used in Proposition~\ref{prop-hetero-conv}, there exists some $N$ independent constant $C$ such that
\bea
&&\sup_{t\in[0,T]}\mbb{E}\Bigl[|\vp_t^{\rm He}-\vp_t^{\rm MFG}|^2\Bigr]+
\mbb{E}\Bigl[\sup_{t\in[0,T]}\bigl|\mbb{E}[\vp_t^{\rm He}|\ol{\calf}_t^0]-\vp_t^{\rm MFG}\bigr|^2\Bigr]\nn \\
&&\leq C\Bigl(\sup_{t\in[0,T]}\mbb{E}\Bigl[W_2(\ol{\mu}_t^N,\call^0_t(y_t))^2\Bigr]+\mbb{E}\Bigl[W_2(\ol{\mu}_g^N, \call_g^0)^2\Bigr]\Bigr)\nn \\
&&\qquad+C\frac{1}{N}\sum_{i=1}^N \mbb{E}\Bigl[|\del {G}_i|^2+\int_0^T 
\bigl(|\del F_i(t)|^2+|\del B_i(t)|^2+|\del \sigma_i^0(t)|^2+|\del \sigma_i(t)|^2\bigr)dt\Bigr]~. \nn
\eea
\begin{proof}
The desired estimate follows from Proposition~\ref{prop-hetero-conv}.
It is easy to check
\be
\begin{split}
|\vp_t^{\rm He}-\vp_t^{\rm MFG}|^2&\leq 2\Bigl|\frac{1}{N}\sum_{i=1}^N (\ch{Y}_t^i-y^i_t)\Bigr|^2+2|m(\ol{\mu}_t^N)-\vp_t^{\rm MFG}|^2  \\
&\leq  2\frac{1}{N}\sum_{i=1}^N |\Del \ch{Y}_t^i|^2+2W_2(\ol{\mu}_t^N, \call_t^0(y_t))^2, 
\end{split}\nn
\ee
which gives the estimate for the first term. 

Using the fact that $\mbb{E}[y_t^1|\ol{\calf}_t^0]=\mbb{E}[y_t^i|\ol{\calf}_t^0]$ for any $i\geq 1$, 
we have
\be
\begin{split}
\mbb{E}\Bigl[\sup_{t\in[0,T]}\bigl|\mbb{E}[\vp_t^{\rm He}|\ol{\calf}_t^0]-\vp_t^{\rm MFG}\bigr|^2\Bigr] 
&=\mbb{E}\Bigl[\sup_{t\in[0,T]}\Bigl| \mbb{E}\bigl[\frac{1}{N}\sum_{i=1}^N (\ch{Y}_t^i-y_t^i)|\ol{\calf}_t^0\bigr]\Bigr|^2\Bigr] \\&
\leq  \mbb{E}\Bigl[\mbb{E}\Bigl[\sup_{t\in[0,T]}|\Del \ch{Y}_t^i|^2|\ol{\calf}^0\Bigr]\Bigr]=\mbb{E}\Bigl[\sup_{t\in[0,T]}|\Del \ch{Y}_t^i|^2\Bigr]~.
\end{split} \nn
\ee
This gives the estimate for the second term.
\end{proof}
\end{corollary}

\begin{remark}
Corollary~\ref{co-price-stability} implies that the market clearing price converges to the mean-field limit $\vp^{\rm MFG}$
if the difference of coefficients functions $(\del G_i, \del F_i, \del B_i,\del \sigma_i^0,\del \sigma_i)_{i\geq 1}$
converges to zero in the large population limit $N\rightarrow \infty$. 
It is clear that any deviation from the limit coefficient functions $(\ol{g}, \ol{f}, l, \sigma^0, \sigma)$
among the finite number of agents does not affect this convergence.
\end{remark}

\section{Conclusions and Discussion}
In this work, we prove the existence of a unique market clearing equilibrium among the
heterogeneous agents of finite population size under the assumption that they are the price takers.
We show the strong convergence to the corresponding mean-field limit given in \cite{Fujii-Takahashi}
under appropriate conditions. In particular, we provide the stability relation between the market clearing price for the heterogeneous agents
and that for the homogeneous mean-field limit.
An extension to multiple populations~\cite{Fujii-mfg} as studied in Section 6 of \cite{Fujii-Takahashi} looks straightforward.
In the work \cite{Fujii-Takahashi-major}, we have studied the similar problems 
in the presence of a major agent, who has a non-negligible market share and hence 
receives a direct price impact from his/her trading.

One of the important remaining issues is to develop numerical evaluation technique so that we can analyze
the dynamics of equilibrium price. In particular, understanding the change of volatility of the price process with respect to those
of risk-averseness of agents and the order flows from the OTC clients will provide us an important insight of 
securities markets. Adopting the linear-quadratic setup may provide us  a semi-analytic solution for this problem.

\subsection*{Acknowledgements}
The authors thank for anonymous referees for valuable comments.
One of the authors (M.F.) also thanks professors Dena Firoozi and Tomoyuki Ichiba 
for their valuable comments at SIAM Annual Meeting (AN21).



\end{document}

%% file: Fmacro-2015.tex

\newtheorem{definition}{Definition}[section]
\newtheorem{assumption}{Assumption}[section]
\newtheorem{condition}{$[$ C}
\newtheorem{lemma}{Lemma}[section]
\newtheorem{proposition}{Proposition}[section]
\newtheorem{theorem}{Theorem}[section]
\newtheorem{remark}{Remark}[section]
\newtheorem{example}{Example}[section]
\newtheorem{corollary}{Corollary}[section]
\def\n{{\bf n}}
\def\A{{\bf A}}
\def\B{{\bf B}}
\def\C{{\bf C}}
\def\D{{\bf D}}
\def\E{{\bf E}}
\def\F{{\bf F}}
\def\G{{\bf G}}
\def\H{{\bf H}}
\def\I{{\bf I}}
\def\J{{\bf J}}
\def\K{{\bf K}}
\def\L{{\bf L}}
\def\M{{\bf M}}
\def\N{{\bf N}}
\def\O{{\bf O}}
\def\P{{\bf P}}
\def\Q{{\bf Q}}
\def\R{{\bf R}}
\def\S{{\bf S}}
\def\T{{\bf T}}
\def\U{{\bf U}}
\def\V{{\bf V}}
\def\W{{\bf W}}
\def\X{{\bf X}}
\def\Y{{\bf Y}}
\def\Z{{\bf Z}}
\def\cala{{\cal A}}
\def\calb{{\cal B}}
\def\calc{{\cal C}}
\def\cald{{\cal D}}
\def\cale{{\cal E}}
\def\calf{{\cal F}}
\def\calg{{\cal G}}
\def\calh{{\cal H}}
\def\cali{{\cal I}}
\def\calj{{\cal J}}
\def\calk{{\cal K}}
\def\call{{\cal L}}
\def\calm{{\cal M}}
\def\caln{{\cal N}}
\def\calo{{\cal O}}
\def\calp{{\cal P}}
\def\calq{{\cal Q}}
\def\calr{{\cal R}}
\def\cals{{\cal S}}
\def\calt{{\cal T}}
\def\calu{{\cal U}}
\def\calv{{\cal V}}
\def\calw{{\cal W}}
\def\calx{{\cal X}}
\def\caly{{\cal Y}}
\def\calz{{\cal Z}}
%
\def\sskip{\hspace{0.5cm}}
\def\simleq{ \raisebox{-.7ex}{\em $\stackrel{{\textstyle <}}{\sim}$} }
\def\leqsim{ \raisebox{-.7ex}{\em $\stackrel{{\textstyle <}}{\sim}$} }
\def\ep{\epsilon}
\def\half{\frac{1}{2}}
\def\iku{\rightarrow}
\def\Iku{\Rightarrow}
\def\ikup{\rightarrow^{p}}
\def\inclusion{\hookrightarrow}
\def\cadlag{c\`adl\`ag\ }
\def\up{\uparrow}
\def\down{\downarrow}
\def\doti{\Leftrightarrow}
\def\douti{\Leftrightarrow}
\def\dochi{\Leftrightarrow}
\def\douchi{\Leftrightarrow}%
\def\yy{\\ && \nonumber \\}
\def\y{\vspace*{3mm}\\}
\def\nn{\nonumber}
\def\be{\begin{equation}}
\def\ee{\end{equation}}
\def\bea{\begin{eqnarray}}
\def\eea{\end{eqnarray}}
\def\beas{\begin{eqnarray*}}
\def\eeas{\end{eqnarray*}}
%
\def\hd{\hat{D}}
\def\hv{\hat{V}}
\def\hsd{{\hat{d}}}
\def\hx{\hat{X}}
\def\hsx{\hat{x}}
\def\bsx{\bar{x}}
\def\bsd{{\bar{d}}}
\def\bx{\bar{X}}
\def\ba{\bar{A}}
\def\bb{\bar{B}}
\def\bc{\bar{C}}
\def\bv{\bar{V}}
\def\balpha{\bar{\alpha}}
\def\bbalpha{\bar{\bar{\alpha}}}
\def\combi{\l(\begin{array}{c}\alpha\\ \beta \end{array}\r)}
\def\f{^{(1)}}
\def\s{^{(2)}}
\def\ss{^{(2)*}}
\def\l{\left}
\def\r{\right}
\def\a{\alpha}
\def\b{\beta}
\def\L{\Lambda}